\documentclass[11pt,a4paper]{article}
\usepackage[T1]{fontenc}
\usepackage[utf8]{inputenc}
\usepackage{amsmath,amssymb,bm,bbm}
\usepackage{braket}
\usepackage{graphicx}
\usepackage{enumitem}
\usepackage[numbers,sort&compress]{natbib}
\usepackage{setspace}
\usepackage[margin=25mm]{geometry}
\usepackage[font=small,labelfont=bf,labelsep=period]{caption}
\usepackage{microtype}
\usepackage{pdflscape}
\usepackage{hyperref}
\setcitestyle{super,comma}
\setlist[enumerate]{leftmargin=2.5em}
\usepackage{booktabs}
\usepackage{tabularx}
\usepackage{threeparttable}
\usepackage{url}
\usepackage{comment}

\title{\bfseries High-fidelity control of long-lived data qubits via a noisy mediator}
\author{Nguyen H. Le$^{1,\dagger}$, Tikai Chang$^{1,\dagger}$, Eran Ginossar$^{2}$ and Michael Stern$^{3,*}$\\[0.8em]
\small $^{1}$Qolumbus SAS, 78 Boulevard de la République, 92100 Boulogne-Billancourt, France\\
\small $^{2}$School of Mathematics and Physics, University of Surrey, GU2 7XH, United Kingdom\\
\small $^{3}$Department of Physics, Technion--Israel Institute of Technology, Haifa 3200003, Israel\\[0.5em]
\small $^{\dagger}$These authors contributed equally to this work.\\
\small $^{*}$Correspondence: mstern@technion.ac.il}
\date{}

\begin{document}
\maketitle

\begin{abstract}
Long-lived qubits are attractive for storing quantum information, but they are often difficult to control or couple directly. A common strategy is to use mediator qubits to generate effective interactions for gate implementation. However, the mediator typically has a shorter coherence time than the data qubits, so mediator relaxation and dephasing are expected to limit the achievable gate fidelity. Here, we show that this limitation can be overcome. We present a  reset-assisted control method that suppresses mediator-induced errors and enables high-fidelity single- and two-qubit gates on a quantum register of long-lived data qubits. Although mediator quantum jumps render the data-qubit evolution stochastic, repeated mediator reinitialization drives the resulting evolution paths toward the desired target operation. An analytical quantum-trajectory analysis  reveals the physical mechanism of error suppression and provides compact fidelity formulas for adapting the method to different platforms. Our results relax the requirement for long mediator coherence,
broadening the range of mediators that can support high-fidelity
control.
\end{abstract}
\clearpage
\section*{Introduction}
Quantum information processing requires qubits that are sufficiently isolated to preserve coherence, yet sufficiently controllable to allow fast initialization, readout, and entangling operations \cite{DiVincenzo2000,Preskill2018,Krantz2019}.
Mediator-based architectures provide a way to reconcile these competing demands by separating the relevant functionalities among different degrees of freedom. Long-lived data qubits are used to store quantum information, while a more accessible mediator is used to control, read out, or generate effective coupling between the data qubits. Hybrid quantum systems provide a natural  realization of this idea \cite{kurizki_quantum_2015} by using a different qubit type as the mediator. Representative examples include optically addressable NV-center electron spins coupled to nearby nuclear spins \cite{childress_coherent_2006,Dutt2007,Taminiau2014}, silicon quantum dots coupled to individual nuclear spins \cite{Hensen2020}, and more recently, superconducting flux qubits acting as quantum buses for single bismuth donors in silicon \cite{chang2024}.

The separation of functionalities also offers a significant advantage for scaling. Because the data qubits do not need to be driven or measured directly during operation, control and readout resources can be concentrated on the mediator. This can substantially reduce the number of control lines, measurement channels, and local interfaces that would otherwise need to be routed to the data qubits, thereby easing integration constraints in large-scale devices~\cite{jordan2026}.

However, a central challenge is that using the mediator qubit to perform operations on the data qubits inevitably exposes those operations to  mediator dissipation and dephasing \cite{Mundada2019,Chu2021,Ding2023,Singh2026,Douce2015}. The mediator is useful precisely because it is more accessible, but this accessibility often comes with stronger coupling to environmental noise. As a result, gate fidelity is limited by the  shorter coherence time of the mediator. Several strategies have been developed to mitigate errors introduced by dissipative auxiliary systems. Error-transparent gate constructions \cite{vy_error-transparent_2013,kapit_error-transparent_2018,ma_error-transparent_2020} can protect logical operations against selected ancilla errors, but rely on an encoded quantum-error-correcting structure. Path-independent gates reduce sensitivity to the path followed by the auxiliary system, but are generally heralded and require postselection \cite{ma_path-independent_2020}. 


In this paper, we introduce a reset-assisted control method that suppresses errors arising from a general mediator noise model and requires neither encoding nor postselection. By dividing a gate into multiple short segments and reinitializing the mediator in a suitably chosen state at the beginning of each segment, the protocol enables quantum gates that are resilient to mediator relaxation, dephasing, and general Pauli errors. Although mediator noise generates
stochastic data-qubit trajectories, repeated reinitialization causes
these trajectories to converge toward the desired target operation.
The suppression becomes increasingly effective as the number of
segments increases, allowing the mediator-induced contribution to the
infidelity to be reduced below a desired threshold until other
imperfections become dominant. High fidelity  can therefore be
achieved even when the mediator coherence time is comparable or shorter than the
total gate duration.


Gate performance is analyzed for an experimentally motivated architecture in which long-lived impurity-spin  qubits are mediated by a superconducting flux qubit \cite{chang2024}. Fidelity approaching 99.99\% can be obtained for both single-qubit and two-qubit gates, despite the duration of the two-qubit gate exceeding the coherence time of the flux qubit.  We discuss a concrete implementation using a cross-resonance
drive and present a general control framework for systems with a controllable auxiliary degree of freedom. Cross-platform estimates based on representative
experimental parameters for quantum-dot and NV nuclear-spin registers,
fluxonium qubits coupled through a transmon, and Rydberg-mediated
neutral-atom systems further indicate that reset-assisted control can suppress mediator-induced two-qubit gate error to $\simeq 10^{-4}$ across substantially different physical platforms.
\\
\\
\begin{figure}[h]
    \centering
\includegraphics[width=1\linewidth]{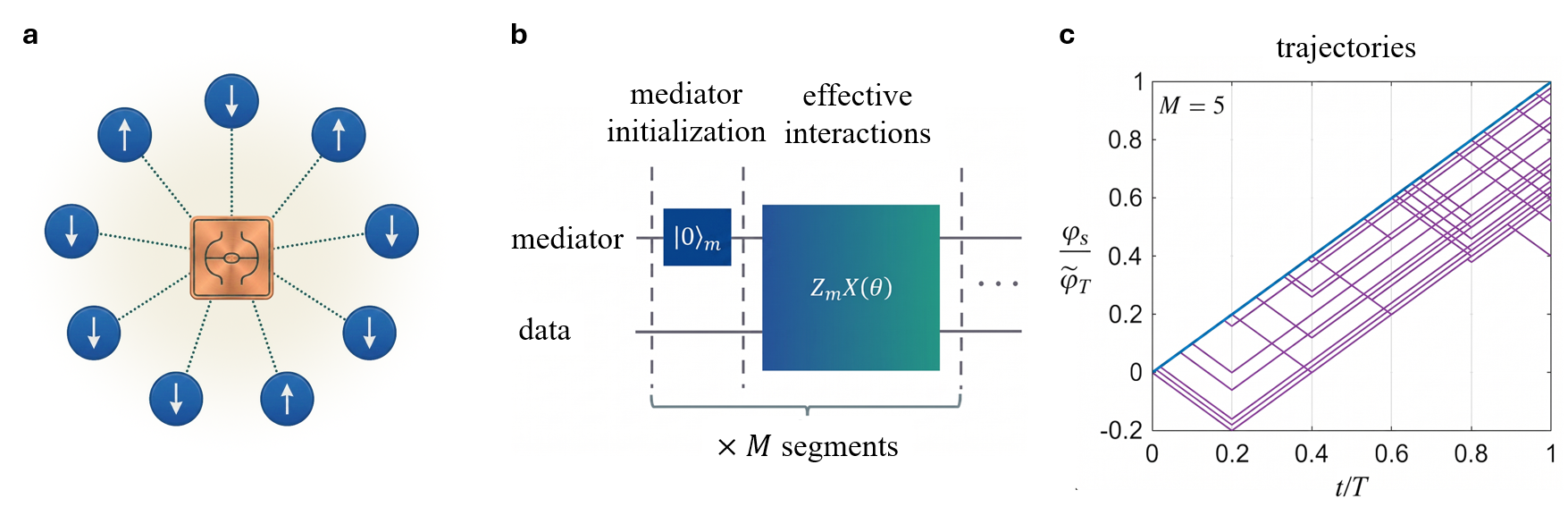}
\caption{\textbf{Architecture and reset-assisted single-qubit control.} (\textbf{a}) Schematic of a processor in which multiple data qubits are coupled to a common mediator qubit. (\textbf{b}) A control segment consists of mediator initialization followed by an effective interaction implementing a single-qubit $X$ rotation. (\textbf{c}) Rotation angle during a $\pi/2$ $X$ rotation for 500 randomly generated trajectories with $M=5$. Error-free trajectories (blue) follow a straight line from 0 to 1. Reinitializing the mediator after each interval of duration $0.2T$ refocuses trajectories displaced by initialization errors and mediator quantum jumps (purple).}
    \label{fig:gate}
\end{figure}
\section*{Results}
\subsection*{Reset-assisted single-qubit gates}
Figure~\ref{fig:gate}a illustrates a system of $n$ long-lived data qubits coupled to a central mediator qubit through tunable effective interactions.  By applying a  drive to the mediator, one selectively activates
a two-body effective Hamiltonian between the mediator qubit and a chosen data qubit, for instance the $k$th one, with $k\in\{1,\cdots,n\}$. Although the  method is applicable with any effective Hamiltonian dominated by two-body terms, the example of a $Z\!X$ Hamiltonian is discussed for simplicity,
\begin{equation}
    \tilde{H}_k = \tilde{g}_k \, Z_m X_k,
\end{equation}
where \(\tilde{g}_k\) is the effective interaction strength between the data qubit and the mediator. Specific  drives that achieve an effective $ZX$ interaction are given below and in Supplementary Note 1.

Assuming that the mediator is initialized in the  $\ket{0}_m$ eigenstate of  $Z_m$  and that the data register is in an $n$~-~qubit state $\ket{\psi}$, the evolution under  $\tilde{H}_k$ for a duration $T$ yields
\begin{equation}\label{eq:sqrotation}
e^{-i\tilde{g}_kT Z_mX_k}\ket{0}_m\otimes\ket{\psi}=\!\ket{0}_m \otimes  e^{- i\tilde{g}_kT X_k}\ket{\psi},
\end{equation}
resulting in an effective rotation on the  $k$th data qubit around its $X$ axis, with rotation angle $\tilde{g}_k T$. Alternatively, if the mediator is initialized in $\ket{1}_m$, the rotation  is in the opposite direction.

The factorization of dynamics in equation~\eqref{eq:sqrotation} relies on  the mediator qubit remaining in $\ket{0}_m$ for the whole gate duration, which no longer holds in the presence of mediator  decoherence. We consider  a general mediator-noise model comprising Pauli errors $X_m,Y_m,Z_m$  and relaxation error $L_m\equiv\ket{0}\!_m\bra{1}_m$, with rates $\gamma_x,\gamma_y,\gamma_z$ and $\gamma_L$, respectively. The effects of these errors are described by the dissipator
\begin{equation}\label{eq:linblad}
\tilde{\mathcal{L}}[\rho] = \sum_{\mu=x,y,z,L} \gamma_\mu \left( \Lambda_\mu \rho \, \Lambda_\mu^{\dagger} - \frac{1}{2} \left\{ {\Lambda_\mu^\dagger} \Lambda_\mu, \rho \right\} \right),
\end{equation}
in the quantum master equation, with jump operators $\Lambda_{\mu}\in\{X_m,Y_m,Z_m,L_m\}$. The data qubit’s coherence time when not effectively coupled to the mediator is assumed to be much longer than the gate duration; therefore, data-qubit decoherence channels other than those induced by the noisy mediator are negligible.

To reduce the gate infidelity due to mediator decoherence, the total gate duration $T$ is divided into $M$ segments of small length
$\tau = T/M$, and at the start of each segment, the mediator qubit is reinitialized to $\ket{0}_m$ with a small preparation error  $\zeta$ such that the post-initialization state  is
\begin{equation}
\rho_m^{\text{init}} = (1-\zeta)\lvert 0\rangle_m\langle 0 \rvert_m + \zeta\lvert 1\rangle_m \langle 1 \rvert_m,
\end{equation}
followed by application of the effective interaction $\tilde{H}_k$, as illustrated in Fig.~\ref{fig:gate}b. In general, the density matrix may contain a small off-diagonal term, which has a negligible leading-order effect (Supplementary Note 2). It is now shown that the  rotation of the data qubit follows a random walk with a rotation angle given by a sample average of the small random angles accumulated in each segment. The law of large numbers dictates that the variance of this angle, proportional to the gate infidelity, scales as the inverse of the number of segments, and hence can be reduced arbitrarily  by increasing $M$.

This random walk of the data qubit is best understood in the quantum-trajectory framework \cite{plenio_quantum-jump_1998,molmer_monte_1993}. For a single segment, the data-qubit trajectory is specified by the initial mediator state and by whether an error in the set $\{X_m, Y_m, Z_m,L_m\}$ occurs. When the segment duration $\tau$ and the initialization error $\zeta$ are sufficiently small, trajectories containing two or more error events  have negligible probability and are therefore neglected. If an initialization error occurs, with probability $\zeta$, and the mediator qubit starts in $\ket{1}_m$, the data qubit is rotated in the opposite direction with rotation angle $-\theta$ where $\theta=\tilde{g}_k\tau$. If the mediator qubit starts in $\ket{0}_m$, a relaxation error $L_m$ or phase-flip error $Z_m$ leaves the state of the mediator qubit unchanged, and the corresponding dynamics remains unaffected with rotation angle $\theta$.
However, a bit-flip error, $X_m$ or $Y_m$, occurring at time $t \in [0,\tau]$ flips the state of the mediator qubit to $\ket{1}_m$, and the data qubit is then rotated in the opposite direction for the remaining duration $\tau - t$, resulting in the net rotation angle $\tilde{g}_kt - \tilde{g}_k(\tau - t) = \theta(2t/\tau - 1)$. The probability that either an $X_m$ or $Y_m$ error event occurs between $t$ and $t+dt$ is given by $\gamma\,dt$, where
\begin{equation}
\gamma=\gamma_x+\gamma_y.
\end{equation}

Therefore, for a single trajectory, the unitary evolution of the data qubit can be described by an $X$ rotation with a random rotation angle
\begin{equation}
U_s=\exp\{-i s \theta X_k\},
\end{equation}
where $s$ is a random factor with the following  discrete-continuous distribution, up to first order in $\gamma\tau$ and $\zeta$,
\begin{align*}
s &= - 1, \quad\ \ \ \ \ \text{with probability } \zeta, \\
s &= 2t/\tau - 1, \ \text{with probability } \gamma\,dt, \ \text{for } 0 \le t \le \tau, \\
s &= 1, \quad \quad \quad \  \, \text{with probability } 1 - \zeta-\gamma\tau.
\end{align*}
The mean and variance of $s$ are
\begin{align}
&\langle s \rangle = 1 - 2\zeta-\gamma\tau + \int_0^\tau\!\!\! \gamma\,dt \left(\frac{2t}{\tau} - 1\right) = 1 -2\zeta- \gamma\tau,\nonumber \\
&\langle\delta s^2\rangle = \langle s^2 \rangle - \langle s \rangle^2 = 4\zeta+\frac{4}{3}\gamma\tau.
\end{align}

Over  $M$ segments, a trajectory is  characterized by a set of $M$ independent random factors $\bm{s} = \{s_1, s_2, \dots, s_M\}$, where $s_j$ denotes the random factor associated with the $j$th segment. The rotation on the data qubit  undergoes a random-walk-type evolution. For a given trajectory, the evolution unitary is
\begin{equation}
U_{\bm{s}} = \exp\{-i\varphi_{\bm{s}} X_k\},
\end{equation}
where the total rotation angle
\begin{equation}
\varphi_{\bm{s}} = \theta \sum_{j=1}^M s_j = \tilde{g}_kT \left( \frac{1}{M}\sum_{j=1}^M s_j \right),
\end{equation}
 is determined by the average of the $M$ independent random factors $s_j$. It has mean value
\begin{equation}
\langle \varphi \rangle = \tilde{g}_kT \langle s \rangle = \tilde{g}_kT(1-2\zeta-\gamma\tau),
\end{equation}
indicating that the average rotation rate is reduced due to the errors. Crucially, as a sample average, the rotation angle has a variance that scales as $1/M$ times the variance of $s_j$, according to the law of large numbers,
\begin{equation}
\langle\delta \varphi^2\rangle= (\tilde{g}_kT)^2 \frac{\langle\delta s^2\rangle}{M}
=  (\tilde{g}_kT)^2\left( \frac{4\zeta}{M}+\frac{4\gamma\tau}{3M}\right).
\end{equation}
The probability of a mediator bit-flip error in each segment, $\gamma\tau\equiv\gamma T/M$, also scales as the inverse of the number of segments. Thus, the repeated reinitialization  results in a $1/M^2$ scaling for the mediator-decoherence's contribution to the  rotation-angle variance:
\begin{equation}
\langle\delta \varphi^2\rangle=(\tilde{g}_kT)^2\left( \frac{4\zeta}{M}+\frac{4\gamma T}{3M^2}\right).
\end{equation}

To implement a rotation with a target phase $\varphi_T$,
\begin{equation}
U_{T} = \exp\{-i\varphi_T X_k\},
\end{equation}
the total duration  is chosen such that $\langle \varphi \rangle=\varphi_T$, yielding
\begin{align}\label{eq:T1q}
T &\approx \tilde{\varphi}_T/\tilde{g}_k, \quad \tilde{\varphi}_T=\varphi_T/(1-2\zeta-\gamma \varphi_T/\tilde{g}_kM).
\end{align}
The data-qubit trajectories during a $\pi/2$ $X$  rotation with $\varphi_T=\pi/4$ are illustrated in Fig.~\ref{fig:gate}c for $2\pi\gamma/\tilde{g}_k=0.5$ and $\zeta=0.01$. The rotation angles $\varphi_{\bm{s}}$ of 500 randomly generated trajectories are shown for $M=5$. While most trajectories are error-free and follow the straight line from 0 to the renormalized target angle $\tilde{\varphi}_T$, occurrences of initialization errors and bit-flip errors lead to sideways trajectories branching away from the error-free trajectories. Reinitialization of the mediator qubit after every $0.2T$ interval refocuses these sideways trajectories, resulting in a reduced spread at the final time.

The final state of the  $k$th data qubit  is the weighted statistical mixture of all possible trajectories
\begin{equation*}
\rho_k(T) = \sum_{\bm{s}} P(\bm{s})\, U_{\bm{s}} \rho_k U_{\bm{s}}^\dagger,
\qquad
P(\bm{s})=\prod_{j=1}^M p(s_j),
\end{equation*}
where $\rho_k$ is the initial density matrix. The state fidelity is given by
$
F(\rho_k)=\operatorname{tr}\bigl(\rho_{k}^{T} \rho_k(T)\bigr),
$
where
$
\rho_{k}^{T}= U_T \rho_k  U_T^\dagger
$
is the target state. The gate fidelity $G$ is obtained by averaging this state fidelity  over all Haar-random pure initial states\cite{zyczkowski_average_2005,mele_introduction_2024},
\begin{equation}
G = \mathbb{E}_{\rho_k}[F(\rho_k)] = \sum_{\bm{s}} P(\bm{s})\, G_{\bm{s}},
\end{equation}
where
$
G_{\bm{s}} = \mathbb{E}_{\rho_k}\Bigl[\operatorname{tr}\left(U_{T}\rho_kU_{T}^{\dagger} U_{\bm{s}}\rho_k U_{\bm{s}}^{\dagger}\right)\Bigr]
$
is the gate fidelity associated with an individual trajectory. This averaging can be done analytically to obtain
\begin{equation}\label{eq:infid1q}
1-G
\approx \frac{2}{3}\langle\delta \varphi^2\rangle
=
\frac{8\tilde{\varphi}_T^2}{3}\frac{\zeta}{M}+
\frac{8\tilde{\varphi}_T^3}{9}\frac{\gamma}{\tilde{g}_kM^2}.
\end{equation}
The reset-assited method can be seen as effectively reducing the initialization error by a factor of $M$ and the mediator-qubit decoherence by a factor of $M^2$. The number of segments can be increased until a desired fidelity threshold is reached.

\subsection*{Reset-assisted two-qubit gates}
Two-qubit gates between a selected pair of data qubits  can be realized by  sequentially activating an effective Hamiltonian $\tilde{H}_{k}=\tilde{g}_kV_k$, coupling the mediator to the $k$th data qubit, and $\tilde{H}_{l}=\tilde{g}_lV_l$, coupling the mediator to the $l$th data qubit. These effective Hamiltonians are applied for durations
\begin{equation}\label{eq:tau}
\tau_k=\theta/\tilde{g}_k, \quad \tau_l=\theta/\tilde{g}_l,
\end{equation}
where  $\theta$ is a small rotation angle, in the following four-pulse sequence
\begin{equation}\label{eq:4pulse}
e^{-i\theta V_{k}}e^{-i\theta V_{l}}e^{i\theta V_{k}}e^{i\theta V_{l}}\approx e^{-\theta^2[V_{k},V_{l}]+O(\theta^3)}.
\end{equation}
This sequence simulates an effective three-body interaction generated by the commutator $[V_{k},V_{l}]$, as follows from a standard application of the Baker--Campbell--Hausdorff formula.

The effective $Z\!X$ interaction is again considered. By conjugating the time evolution of this interaction with $\pm \pi/2$ $X_m$ rotations on the mediator qubit, one can generate an effective $Y\!X$ interaction
\begin{equation}\label{eq:conj}
e^{i(\pi/4)X_m}\, e^{-i\theta Z_m X_k}\, e^{-i(\pi/4)X_m}
=
e^{-i\theta Y_m X_k}.
\end{equation}
The two effective interactions
$\tilde{g}_k Y_m X_k$ and $\tilde{g}_l Z_m X_l$ can now be alternated
in the four-pulse sequence of Eq.~\eqref{eq:4pulse} to simulate the effective three-body interaction
\begin{equation}\label{eq:three-body}
e^{-\theta^2[Y_m X_k, Z_m X_l]} = e^{-2i\theta^2 X_m X_k X_l}.
\end{equation}
With the mediator qubit initialized in the $+1$ eigenstate of  $X_m$, denoted by $\ket{+}_m$, this pulse sequence induces the two-qubit rotation $\exp{\{-2i\theta^2 X_k X_l\}}$. Unlike in the single-qubit case discussed above, the mediator is not reinitialized after each $Y_mX_k$ or $Z_mX_l$ pulse. Instead, the full four-pulse sequence must be completed to realize the three-body $X_mX_kX_l$ interaction.
\begin{figure}[t]
    \centering
\includegraphics[width=0.5\linewidth]{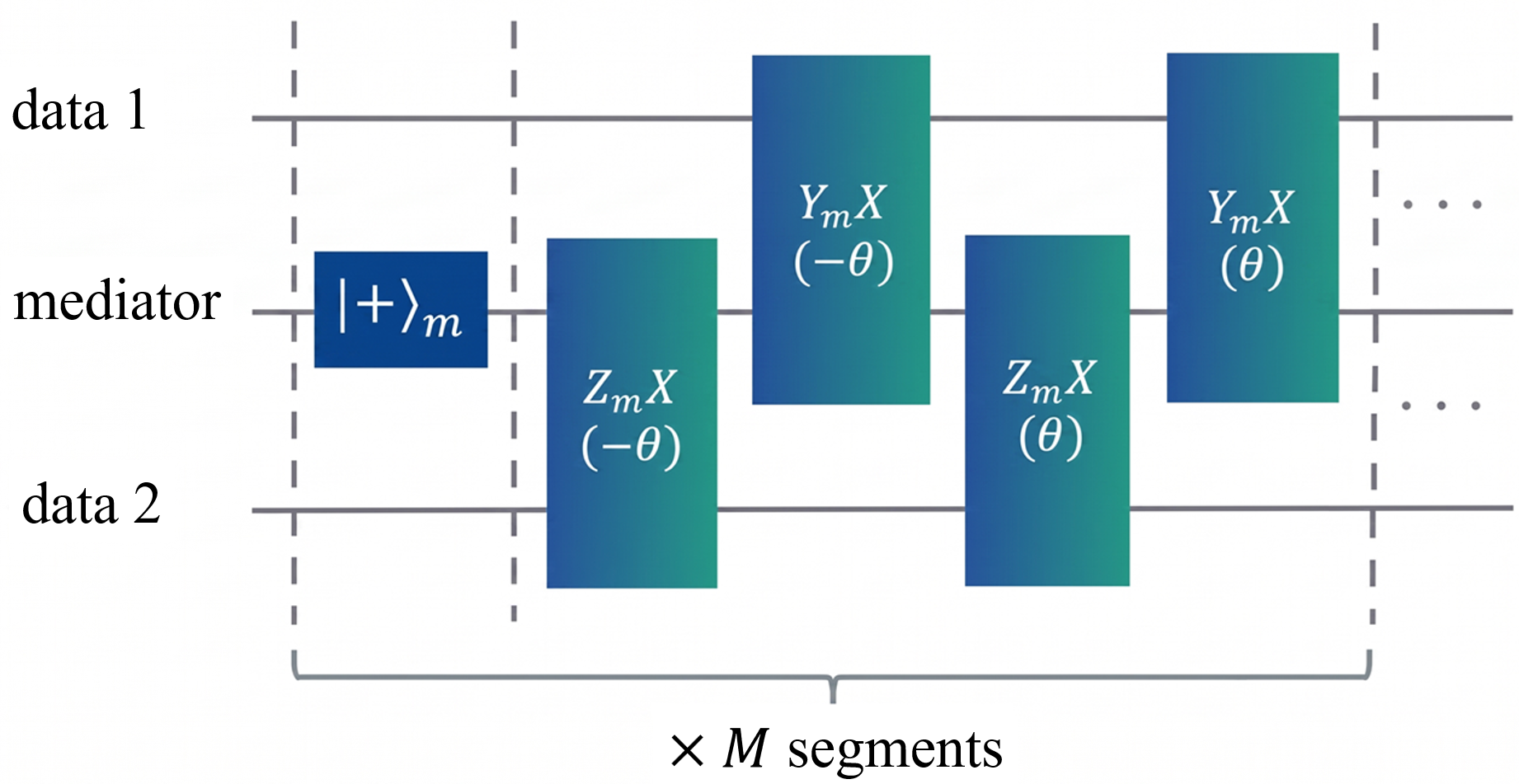}
\caption{\textbf{Reset-assisted two-qubit control sequence.} Each segment consists of mediator initialization followed by the four-pulse sequence that implements a two-qubit $X\!X$ rotation on the data qubits.}
    \label{fig:2q}
\end{figure}

To mitigate mediator decoherence, the total duration is again divided into $M$ segments; the mediator qubit is initialized to $\lvert +\rangle_m$ with preparation fidelity $1-\zeta$ at the start of each segment, followed by application of the four-pulse sequence (Fig.~\ref{fig:2q}).  For a single segment, the trajectory of the data qubits depends on the starting eigenstate of the mediator qubit and on whether an error in the set $\{X_m,Y_m,Z_m,L_m\}$ occurs in the first, second, third, or fourth  pulse. There are 39 distinct classes of trajectories, each resulting in an approximate unitary evolution of the data qubits (Methods).

Over $M$ segments, the unitary evolution associated with an individual trajectory is given by
\begin{align}
&U_{\bm{q},\bm{r},\bm{s}}=e^{i\alpha_{\bm{q}} X_k}e^{i\beta_{\bm{r}} X_l}e^{-i\varphi_{\bm{s}}X_kX_l}, \\
&\alpha_{\bm{q}}=2\theta\sum_{j=1}^M q_j, \quad \beta_{\bm{r}}=2\theta\sum_{j=1}^M r_j,\quad \varphi_{\bm{s}}=2\theta^2\sum_{j=1}^M s_j,\nonumber
\end{align}
where $q_j$, $r_j$, and $s_j$ are random factors in the $j$th segment. Importantly, there are now single-qubit fluctuations determined by $q_j$ and $r_j$, in addition to the fluctuation $s_j$ in the two-qubit rotation angle. While a mediator quantum jump in the single-qubit gate merely reverses the remaining rotation within one segment, a jump in this case breaks the closed loop of the four-pulse sequence and leaves an uncanceled first-order single-qubit kick.

The means and variances  satisfy
\begin{align}\label{eq:var2q}
&\langle \alpha\rangle=\langle \beta\rangle=0, \quad \langle\delta\alpha^2\rangle+\langle \delta\beta^2\rangle\approx4M\theta^2\Gamma (\tau_k+\tau_l),
\nonumber \\
& \langle \varphi\rangle\approx2M\theta^2(1-2\zeta),\quad \langle\delta\varphi^2\rangle\approx16M\theta^4 \zeta.
\end{align}
 where $\Gamma=
(5/3)\gamma_x+(2/3)\gamma_y+\gamma_z+(7/12)\gamma_L
$ (Supplementary Note 3).

For a target rotation phase  $\varphi_T$, $\theta$ and $M$ are chosen such that $\langle \varphi\rangle =\varphi_T$, or
\begin{equation}\label{eq:theta2q}
\theta=\sqrt{\tilde{\varphi}_T/2M},\qquad
\tilde{\varphi}_T=\varphi_T/(1-2\zeta).
\end{equation}
Therefore, after substituting $\theta$ from equation~\eqref{eq:theta2q} into equation~\eqref{eq:tau}, the gate duration is
\begin{equation}\label{eq:T2q}
T=2M(\tau_k+\tau_l)=2\sqrt{2M\tilde{\varphi}_T}/\bar{g},
\end{equation}
where $\bar{g}$ is the harmonic mean of $\tilde{g}_k$ and $\tilde{g}_l$.

The gate infidelity is obtained by averaging over all single-trajectory fidelities, yielding
$
1-G=(4/5)\left[\langle \delta\alpha^2\rangle+\langle\delta\beta^2\rangle+\langle\delta \varphi^2\rangle\right]
$,
and after substituting $\tau_{k,l}$ from equation~\eqref{eq:tau} and $\theta$ from Eq.~\eqref{eq:theta2q} into Eq.~\eqref{eq:var2q},
\begin{align}\label{eq:infid2q}
1-G&=
\frac{8 \tilde{\varphi}_T^{3/2}}{5}\frac{\Gamma}{\bar{g}}\sqrt{\frac{2}{M}}
+\frac{16\tilde{\varphi}_T^2}{5}\frac{\zeta}{M}.
\end{align}
Compared with the single-qubit gate, the mediator-decoherence contribution is suppressed more slowly, scaling as $1/\sqrt{M}$
 rather than $1/M^2$, while the initialization-error contribution remains proportional to $1/M$.

\subsection*{Gate performance}
\begin{figure}[t]
    \centering
\includegraphics[width=0.5\linewidth]{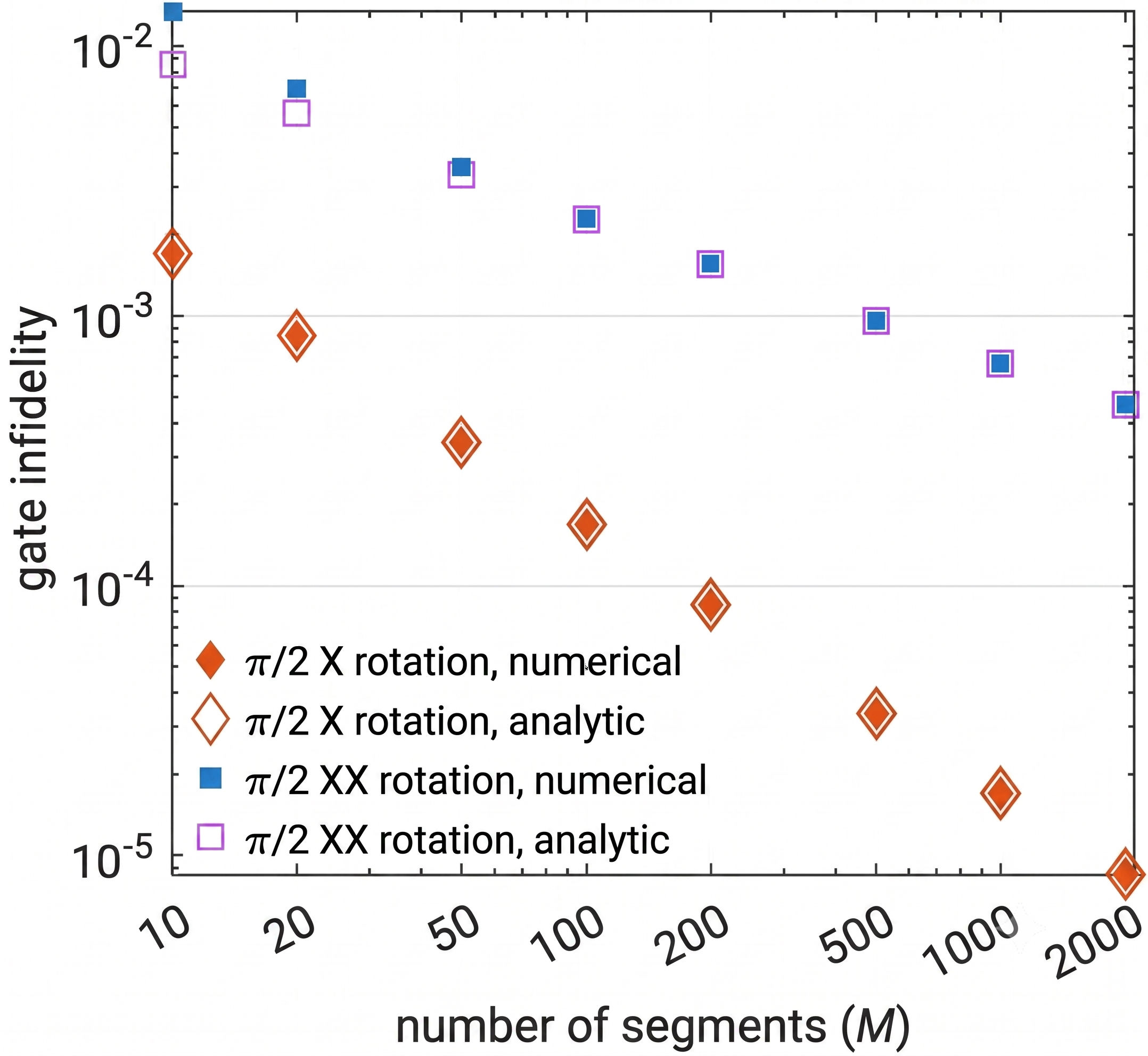}
\caption{\textbf{Suppression of mediator-induced gate errors.} Gate infidelity as a function of the number of segments $M$ for a single-qubit $\pi/2$ $X$ rotation (diamonds) and a two-qubit $\pi/2$ $X\!X$ rotation (squares). Filled symbols show numerical master-equation simulations; open symbols show the analytic predictions.}
    \label{fig:infid}
\end{figure}

Figure~\ref{fig:infid} illustrates the  analytic gate infidelity, obtained from Eqs.~\eqref{eq:infid1q} and \eqref{eq:infid2q}, and the numerically simulated gate infidelity, obtained by propagating a master equation with a Lindbladian describing the error channels, for the single-qubit $\pi/2$ $X$ rotation and two-qubit $\pi/2$ $X\!X$ rotation  (with $\varphi_T = \pi/4$) of the data qubits. The simulation parameters correspond to an  implementation of a processor using bismuth donors  mediated by a flux qubit with relaxation time $T_1=100$~$\mu \rm{s}$ and pure dephasing time $T_{\phi}=20$~$\mu \rm{s}$ \cite{Chang2022}, assuming  $\tilde{g}=2\pi\times 100$~$\rm{kHz}$. The flux qubit is reinitialized with a preparation error $\zeta=0.01$. A  cross-resonance drive applied to the mediator realizes the effective $Z\!X$ interaction, as described in more detail in the next section, and maps mediator relaxation and dephasing into the full error set  $\{X_m,Y_m,Z_m,L_m\}$ with rates  given  in terms of $T_1$ and $T_{\phi}$. The results in Fig.~\ref{fig:infid} are obtained for a drive with an amplitude-detuning ratio ${\cal E}/\Delta=10$, where ${\cal E}$ is the amplitude and $\Delta$ the detuning between the spin and the flux qubit.

There is a remarkably strong agreement between the analytic and numerical results. The small deviation in the two-qubit case, noticeable at $M\leq 20$, is due to higher-order  trajectories, where two or more jump events occur, which are neglected in the analytic estimate. As shown, gate infidelity decreases as the number of segments increases. For the considered parameter values, the single-qubit gate infidelity is dominated by the initialization error  contribution, while the two-qubit gate infidelity is dominated by the mediator-decoherence-induced  errors.

Single-qubit gate infidelity $\sim 10^{-4}$  and two-qubit gate infidelity $\sim 10^{-3}$ are achieved at $M=100$, for which the gate durations, given by Eqs.~\eqref{eq:T1q} and \eqref{eq:T2q}, are $1.2$~$\mu \rm{s}$ and $40$~$\mu\rm{s}$, respectively. Increasing $M$  to $1000$, the two-qubit gate infidelity approaches $\approx 6\times 10^{-4}$ even when the gate duration, $\approx 120 \ \mu\rm{s}$, exceeds the  flux-qubit relaxation time. Further reductions in infidelity can be achieved by increasing $M$, at the cost of a longer gate duration. A Si:Bi spin  operated at a clock transition has a coherence time approaching $\sim 10$~$\rm{s}$ \cite{bienfait_2016,wolfowicz_2013}. Thus, spin decoherence not induced by the mediator flux qubit can be neglected over these gate durations.

\subsection*{Cross-resonance implementation}
The effective $Z\!X$ Hamiltonian can be realized with a native $X\!X$ interaction and a cross-resonance drive \cite{Rigetti2010,Chow2011,Malekakhlagh2020}
$
H_d(t) = {\cal E} X_m \cos \omega_k t ,
$
where $\omega_k$ is the resonance frequency of the targeted $k$th data qubit. The system Hamiltonian is
\begin{equation}
H(t) = -\frac{\omega_m}{2} Z_m - \frac{\omega_k}{2} Z_k + g_k X_m X_k + H_d(t),
\end{equation}
where $g_k$ is the strength of the native $X\!X$ interaction between the mediator and the $k$th data qubit. The mediator qubit also undergoes relaxation with rate $\gamma_1=1/T_1$ and is subject to $1/f$ noise leading to pure dephasing with rate $\gamma_2=1/2T_{\phi}$.

Let $\Delta\! =\! \omega_k - \omega_m$ be the data-mediator detuning. There exists a doubly-rotating frame defined by
\begin{equation}
\exp\left\{i\frac{\omega t}{2}(Z_m+Z)\right\}\exp{\left\{-i\frac{t}{2}\left({\Delta Z_m+{\cal E}X_m}\right)\right\}}
\end{equation}
that transforms the Hamiltonian to (Supplementary Note 1)
\begin{equation}\label{eq:ZX}
\tilde{H} \approx (g_k/2)\sin{\eta}\,  Z_m' X_k,
\end{equation}
where $Z'_m$ is an operator in the tilted frame defined by
\begin{align}
Z_m'\! &= \! Z_m\cos{\eta}  +  X_m\sin{\eta},
\end{align}
with $\eta=\arctan({\cal E}/\Delta)$.  Shifting the drive phase by an angle $\phi$ yields the effective $Z'_m(\cos{\phi}X_k+\sin{\phi}Y_k)$ interaction for implementing a  rotation  along any axis in the $X\!Y$ plane.

In the doubly rotating frame, mediator relaxation and pure dephasing  are transformed into an error channel with jump operators  \begin{align}
&X'_m\equiv X_m\cos{\eta}  -  Z_m \sin{\eta}, \quad Y'_m\equiv Y_m, \nonumber \\
&Z'_m, \quad L'_m\equiv (X'_m+iY'_m)/2,
\end{align}
and respective rates
\begin{align}
\gamma_x&=\gamma_y= (\gamma_1/8)(1-\cos\eta)^2,\nonumber \\
\gamma_z&= (\gamma_1/4)(\sin\eta)^2 + \gamma_2 \cos\eta,\nonumber \\
\gamma_L&= \gamma_1 \cos{\eta},
\end{align}
which is the error model discussed above. The low-frequency $1/f$
contribution to pure dephasing is  filtered by the drive~\cite{Yan2013},
with its effective rate reduced by
$\cos\eta=\Delta/\sqrt{\Delta^2+{\cal E}^2}$. Thus, the strong-drive
choice ${\cal E}/\Delta=10$ used in Fig.~\ref{fig:infid} suppresses
this contribution by approximately one order of magnitude
(Supplementary Note~1).

To initialize the mediator qubit in an eigenstate of $Z'_m$, it suffices to initialize it in an eigenstate of $Z_m$ and either apply  a $Y_m$ rotation with angle $\eta$ or ramp up  the drive amplitude ${\cal E}$ adiabatically. When strong coupling to a resonator is available, fast initialization to the ground state can be achieved by shifting the mediator frequency toward resonance with a resonator mode to facilitate excitation exchange. For mediator flux qubits, initialization can be completed in $\sim 10$-$20$ns (Supplementary Note 4). The total initialization time for $M=100$ is thus comparable to the active-interaction duration of the single-qubit gate, while remaining only $\sim5\%$ of the duration of the two-qubit gate.

Ideally, the data qubits in the register should be spectrally separated on the scale of their coupling strengths to the mediator qubit. However, when there is a spectator with frequency $\omega_s$ close to the frequency $\omega_k$ of the targeted data qubit, with $\omega_s-\omega_k\sim g_k$ and $g_s\sim g_k$, it can be shown that there is no backaction from this spectator qubit on the evolution of the $k$th data qubit in the single-qubit gate, and hence the single-qubit gate infidelity is unaffected.  For the two-qubit gate, the presence of a spectator qubit close to one of the data qubits leads to a moderate increase in the gate infidelity, but this becomes negligible in the limit of large $M$ (Supplementary Note 5).

The effective $Z'_m X_k$ Hamiltonian is derived within the rotating-wave approximation. The effect of rapidly oscillating counter-rotating terms can be effectively canceled when the duration of each segment is chosen to be an integer multiple of their oscillation periods (Supplementary Note 1). In addition, typical fluctuations in the mediator frequency, drive amplitude, and interaction strength result in  additional gate errors below $10^{-6}$ (Supplementary Note 5). The present method is thus robust against various sources of experimental imperfection.

\subsection*{General control framework}
We now describe the reset-assisted control method in its most general setting, in which a short-lived auxiliary quantum system mediates interactions between multiple long-lived data qubits. This auxiliary system may be a single qubit, a collection of qubits, a multi-level qudit, or a bosonic mode. We assume that control Hamiltonians act exclusively on the mediator and that the system has native mediator-data interactions, but no direct interactions between the data qubits. Therefore, entangling operations between data qubits must be generated indirectly through their interactions with the mediator. The general system Hamiltonian in the rotating frame of the noninteracting free terms is
\begin{equation}
{\cal H}(t)=\sum_{i=1}^{m_c} c_i(t) H_m^{(i)}+\sum_{k=1}^n  V_{md}^{(k)}(t),
\end{equation}
where $H_m^{(i)}$ are the control operators acting on the mediator, $c_i(t)$ the control functions, $m_c$ the number of control channels, and $V_{md}^{(k)}(t)$ the native interactions between the mediator and the $k$th data qubit which may become time-dependent in the rotating frame.

The control functions $c_i(t)$ and  control durations are chosen such that the unitary evolution generated by ${\cal H}(t)$ produces a propagator of the form
\begin{equation}
{\cal U}=e^{-i\varphi A_m A},
\end{equation}
where $A_m A$ is a separable effective Hamiltonian that can be expressed as the tensor product of an operator $A_m$ acting on the mediator and, in general, a multi-qubit operator $A$ acting on a selected subset of the data qubits. The set of realizable effective Hamiltonians is given by the dynamical Lie algebra generated by the interaction and control operators in ${\cal H}(t)$ \cite{Albertini2002,dalessandro2021,Mode2024}. For typical two-body interactions and one-body controls, this dynamical Lie algebra contains numerous elements with the required mediator-data separable form \cite{Mode2024}.

By initializing the mediator in an eigenstate of $A_m$, denoted by $\ket{a}_m$ satisfying \begin{equation}
A_m\ket{a}_m=a\ket{a}_m,
\end{equation}
and then applying the mediator drive to implement the propagator ${\cal U}$ on the total system, the multi-qubit gate
\begin{equation}
U_d=e^{-i\varphi a A},
\end{equation}
is implemented on the data qubits. This follows from
\begin{equation}
e^{-i\varphi A_mA}\ket{a}_m\otimes \ket{\psi}=\ket{a}_m\otimes e^{-i\varphi a A}\ket{\psi},
\end{equation}
where $\ket{\psi}$ is an arbitrary state of the data qubits.

In practice, however, acquiring a target phase may require a duration longer than the mediator relaxation or dephasing time. To suppress gate errors induced by mediator decoherence, the following  segmented control protocol can be used:
\begin{enumerate}[label=(\roman*)]
    \item Divide the gate into $M$ short segments.
    \item Initialize the mediator in an eigenstate of the mediator operator $A_m$.
    \item{Apply  the mediator controls with  appropriate shapes and durations to realize the propagator  $e^{-i\delta \varphi A_mA}$ with a small rotation phase $\delta \varphi$}.
    \item{Repeat steps (ii) and (iii) for all $M$ segments, reinitializing the mediator in the same eigenstate at the beginning of each segment.}
\end{enumerate}
The duration of each segment and the number of segments are chosen so that the desired total  phase is obtained.  By increasing the number of segments, the gate error induced by mediator decoherence is reduced until a target infidelity is reached. The required control functions for realizing a target many-body effective Hamiltonian can be obtained using  Floquet engineering \cite{Geier2021,Petiziol2021}  or more general optimal control techniques \cite{Mode2024,Jensen2021,goodwin_adaptive_2022}.

\section*{Discussion}
A proof-of-principle demonstration of reset-assisted control appears within immediate reach in hybrid spin--superconducting architectures. For impurity spins mediated by flux qubits, the key experimental capabilities including strong spin--flux-qubit coupling~\cite{chang2024} and fast measurement-based initialization of flux qubits~\cite{Grover2020}, have already been demonstrated. Integrating multiple flux qubits into a common cavity could additionally provide long-range connectivity between the mediators, thereby enabling mediated interactions between distant impurity spins. Reset-assisted long-range entangling gates in such an architecture would be particularly attractive for implementing quantum error correction with QLDPC codes~\cite{Breuckmann2021}, for which nonlocal parity checks or long-range connectivity can substantially reduce routing overhead.

Beyond this specific implementation, the method applies whenever long-lived data qubits interact through an auxiliary degree of freedom that can be rapidly reinitialized. Table~\ref{tab:cross-platform} summarizes representative experimental parameters for four additional platforms of current interest and estimates the mediator-relaxation contribution to the infidelity of a reset-assisted two-qubit gate. The parameter conversions and underlying assumptions are detailed in Supplementary Note~6. The estimates use $M=100$, corresponding to a tenfold suppression of the leading $M^{-1/2}$ contribution from mediator relaxation. They neglect mediator pure dephasing, which can be filtered by strong driving; initialization errors, which decrease more rapidly as $M^{-1}$, and other platform-specific control imperfections. The estimates therefore isolate the residual gate error arising from mediator relaxation after suppression by reset-assisted control.

\begin{table*}[t]
\centering
\begin{threeparttable}
\caption{
Representative parameters and estimated performance of reset-assisted
two-qubit gates in four additional mediator-based platforms.
The estimates assume $M=100$, $\varphi_T=\pi/4$, and that the gate error is
dominated by mediator relaxation, corresponding to a tenfold suppression
of the leading $M^{-1/2}$ contribution.
The total duration is
$T_{\mathrm{tot}}=T_{\mathrm{active}}+(M-1)t_{\mathrm{reset}}$.
The estimated infidelity contains only the mediator-relaxation
contribution; data-qubit decoherence, reset errors, and platform-specific
control imperfections are not considered.
}
\label{tab:cross-platform}

\small
\setlength{\tabcolsep}{3.5pt}
\renewcommand{\arraystretch}{1.18}

\begin{tabularx}{\textwidth}{@{}Xcccccc@{}}
\toprule
Data qubits / mediator
&
$\bar g/2\pi$
&
$T_{1}$
&
$t_{\mathrm{reset}}$
&
$T_{\mathrm{active}}$
&
$T_{\mathrm{tot}}$
&
$(1-G)_{\mathrm{med}}$
\\
\midrule

\shortstack[l]{
$^{29}\mathrm{Si}$ nuclear spins /\\
quantum-dot electron~\cite{Hensen2020}
}
&
$64.2~\mathrm{kHz}$~\cite{Hensen2020}
&
$0.88~\mathrm{ms}^{\dagger}$~\cite{Zhang2020}
&
$6.3~\mathrm{ns}^{\ast,\dagger}$~
\cite{HarveyCollard2022}
&
$62.2~\mu\mathrm{s}$
&
$62.8~\mu\mathrm{s}$
&
$2.40\times10^{-4}$
\\

\shortstack[l]{
$^{13}\mathrm{C}$ nuclear spins /\\
NV electron~\cite{Taminiau2014}
}
&
$16.5~\mathrm{kHz}$~\cite{Taminiau2014}
&
$3.24~\mathrm{ms}$~\cite{Taminiau2014}
&
$2~\mu\mathrm{s}$~\cite{Taminiau2014}
&
$242~\mu\mathrm{s}$
&
$440~\mu\mathrm{s}$
&
$2.54\times10^{-4}$
\\

\shortstack[l]{
Fluxonium qubits /\\
transmon coupler~\cite{Singh2026}
}
&
$36.0~\mathrm{MHz}^{\ast}$~\cite{Singh2026}
&
$1.2~\mu\mathrm{s}$~\cite{Singh2026}
&
$22~\mathrm{ns}^{\dagger}$~\cite{Chen2024}
&
$111~\mathrm{ns}$
&
$2.29~\mu\mathrm{s}$
&
$3.14\times10^{-4}$
\\

\shortstack[l]{
Ground-state $^{87}\mathrm{Rb}$ atom/\\
 $24D_{5/2}$ Rydberg state
}
&
$21.8~\mathrm{MHz}^{\ast}$~\cite{MacLennan2019}
&
$13.5~\mu\mathrm{s}^{\dagger}$~\cite{Branden2010}
&
$55~\mathrm{ns}^{\ast,\dagger}$~
\cite{Piotrowicz2011,Gutterres2002}
&
$183~\mathrm{ns}$
&
$5.63~\mu\mathrm{s}$
&
$4.61\times10^{-5}$
\\
\bottomrule
\end{tabularx}

\begin{tablenotes}[flushleft]
\footnotesize
\item[] $^{\ast}$ Inferred from reported experimental parameters.
\item[] $^{\dagger}$ Taken from a closely related device or architecture.
\end{tablenotes}

\end{threeparttable}
\end{table*}

The solid-state spin systems in Table~\ref{tab:cross-platform} provide particularly natural extensions of the hybrid architecture considered above. In both cases, long-lived nuclear spins are controlled through a more accessible electronic-spin mediator. For $^{29}\mathrm{Si}$ nuclear spins coupled to a quantum-dot electron, the comparatively strong hyperfine interaction and a projected resonator-assisted reset yield a total gate duration of approximately $63~\mu\mathrm{s}$ and a mediator-induced infidelity of $2.4\times10^{-4}$. For $^{13}\mathrm{C}$ nuclear spins coupled to an NV electron, the weaker hyperfine interactions lead to a longer total duration of approximately $440~\mu\mathrm{s}$, while the millisecond electron relaxation time and established optical initialization still give a comparable mediator-induced infidelity of $2.5\times10^{-4}$. These examples illustrate how rapidly resettable electronic spins can mediate high-fidelity operations on nuclear-spin registers even when direct nuclear control is slow or experimentally inconvenient.

A favorable superconducting implementation is provided by long-lived fluxonium qubits coupled through a driven transmon coupler. In this system, the coupler relaxation time is approximately $1.7~\mu\mathrm{s}$, substantially shorter than the relaxation times of the fluxonium data qubits, and coupler relaxation has been identified as a limiting mechanism for driven two-qubit gates~\cite{Singh2026}. Combining the strong effective interaction with the $22~\mathrm{ns}$ resonator-assisted reset demonstrated in a related tunable-coupler architecture~\cite{Chen2024} gives an estimated mediator-induced infidelity of $3.1\times10^{-4}$. Although the active interaction time is only about $111~\mathrm{ns}$, the repeated resets increase the total duration to approximately $2.3~\mu\mathrm{s}$.

Rydberg-mediated interactions provide a feasible implementation in
neutral-atom processors. In particular, a single $^{87}\mathrm{Rb}$
atom can serve as a Rydberg mediator, while nearby
$^{87}\mathrm{Rb}$ atoms encode the data qubits in the long-lived
$5S_{1/2}$ ground-state hyperfine manifold. Scattering of the
mediator's Rydberg electron from a ground-state data atom produces a
hyperfine-state-dependent energy shift and hence an effective
longitudinal mediator--data interaction. From the measured
$F$-dependent binding energies of long-range
$^{87}\mathrm{Rb}_2$ molecules involving the $24D_{5/2}$ state, an
effective Pauli coupling
$\bar g/2\pi\simeq21.8~\mathrm{MHz}$ can be inferred
\cite{MacLennan2019}. An optical-quench reset can be implemented by
stimulated transfer from $24D_{5/2}$ to the short-lived
$5P_{3/2}$ state, followed by spontaneous decay to the ground-state
manifold. Combining an estimated reset time of $55~\mathrm{ns}$
\cite{Piotrowicz2011,Gutterres2002} with the measured
$29D_{5/2}$ lifetime $T_1=13.5~\mu\mathrm{s}$
\cite{Branden2010} gives a total gate duration of
$5.63~\mu\mathrm{s}$ and a residual mediator-relaxation contribution
of $4.6\times10^{-5}$ (Supplementary Note 6). Because these parameters are drawn from different experiments, the resulting values should be regarded as an order-of-magnitude estimate for a proposed reset-assisted experiment rather than a prediction for an existing device. A similar implementation may also be possible using Rydberg-mediated interactions in dopant-atom architectures~\cite{crane2021}.

Extensions beyond the two-qubit construction considered in Table~\ref{tab:cross-platform} can be obtained within the general control framework whenever the dynamical Lie algebra contains suitable separable mediator--data operators. Of particular interest are effective data-qubit operators containing three-body terms, such as
$A\equiv X_kX_{k+1}X_{k+2}$, which can be used to realize a Toffoli gate~\cite{Xu2026}, or sums of two-body terms, such as
$A\equiv\sum_{kl}(X_kX_l+Y_kY_l)$, which can generate parallel two-qubit operations~\cite{Chen2026}. When direct single-qubit controls are also available on the data qubits, they can be combined with mediator control to enlarge the accessible set of effective Hamiltonians.

These examples show that the relevant hardware criterion is not the mediator lifetime alone, but the balance between the mediator--data coupling strength, the mediator relaxation time, and the speed and fidelity of reinitialization. Mediator noise therefore need not be addressed solely by extending mediator coherence. Repeated reinitialization can instead suppress the transfer of mediator noise to the data register and prevent the mediator lifetime from imposing a hard fidelity limit on the operation. This relaxes an important architectural constraint and broadens the range of physical systems that can serve as practical quantum mediators.

\section*{Methods}
\subsection*{Analytical construction of two-qubit trajectories}

The two-qubit trajectory distribution is obtained perturbatively to
first order in $\zeta$ and $\gamma_\mu\tau_{k,l}$ and to second order
in the pulse angle $\theta$. Trajectories containing more than one
error event within a segment are neglected. The mediator rotations
between the $Z_mX_k$ and $Z_mX_l$ evolutions are assumed to be much
faster than these interaction pulses, so decoherence during the
mediator rotations is also neglected.

For each jump operator $\Lambda_\mu$ occurring at time $t$ during one of the
four interaction pulses, the jump operator is inserted into the interrupted
propagator. The resulting error operator is then propagated through
the remainder of the pulse sequence using commutation relations and
the Baker--Campbell--Hausdorff expansion. After applying the resulting
Kraus operator to the initialized mediator state and tracing out the
mediator, the reduced data-qubit channel can be expressed as a mixture
of unitary trajectory branches. For example, an $X_m$ jump during the
leftmost $Z_mX_k$ pulse produces the two branches
\begin{equation}
U_{\pm}(t)
=
e^{\pm 2i\theta(1-t/\tau_k)X_k}
e^{-2i\theta^2X_kX_l},
\end{equation}
each occurring with probability $\gamma_x\,dt/2$. Repeating this
construction for every jump type and  position gives the
trajectory distribution used in the analytical fidelity calculation.
The complete set of trajectories and their probabilities is provided
in Supplementary Note~3.
\subsection*{Open-system numerical simulations}
To estimate the infidelity shown in Fig.~\ref{fig:infid}, we simulate the Lindblad master equation for the effective $Z_mX_k$ Hamiltonian with dissipator
\begin{equation}\label{eq:linblad2}
\tilde{\mathcal L}[\rho]
=
\sum_{\mu=x,y,z,L}\gamma_\mu
\left(
\Lambda_\mu\rho \Lambda_\mu^\dagger
-
\frac12\left\{\Lambda_\mu^\dagger \Lambda_\mu,\rho\right\}
\right),
\end{equation}
where $\Lambda_\mu\in\{X_m,Y_m,Z_m,L_m\}$ are the mediator jump operators and $\gamma_\mu$ are the corresponding rates.

The simulation begins from $\rho_m^{\rm init}\otimes\ket\psi\!\bra\psi$. The full mediator-data state is propagated through one segment, and the reduced data-qubit state $\rho_d(\tau)$ is obtained by tracing out the mediator. Before the next segment, the total state is reset to $\rho_m^{\rm init}\otimes\rho_d(\tau)$. After all segments have been applied,  the state infidelity is computed between the final reduced data-qubit state and the target state.

The gate infidelity is obtained by averaging the state infidelity over Clifford states. For the single-qubit gate, these are the states reachable from $\ket0$ by single-qubit Clifford operations. For the two-qubit gate, they are the states reachable from $\ket{00}$ by two-qubit Clifford operations.

\section*{Data availability}
The data that support the findings of this work are openly available at \cite{zenodo2}.

\section*{Code availability}
The code used to generate the results is openly available with the supporting data at the repository cited in the Data availability statement.

\section*{Acknowledgements}
M.S. acknowledges support from the Israeli Science Foundation under grant number Quantum ISF 2471/24 and from the Helen Diller Quantum Center at the Technion. E.G. acknowledges support from the Engineering and Physical Sciences Research Council grant number EP/Z000505/1.

\section*{Author contributions}
N.H.L. and T.C. carried out the analytical derivations and numerical simulations. M.S. and T.C. conceptualized the project. All authors discussed the results and contributed to writing the manuscript.

\section*{Competing interests}
The authors declare no competing interests.

\section*{Additional information}
Correspondence and requests for materials should be addressed to Michael Stern.

\clearpage
\onehalfspacing
\hypersetup{hidelinks}
\setcounter{equation}{0}
\setcounter{figure}{0}
\setcounter{table}{0}
\renewcommand{\theequation}{S\arabic{equation}}
\renewcommand{\thefigure}{S\arabic{figure}}
\renewcommand{\thetable}{S\arabic{table}}
\renewcommand{\theHequation}{S.\arabic{equation}}
\renewcommand{\theHfigure}{S.\arabic{figure}}
\renewcommand{\theHtable}{S.\arabic{table}}
\pdfbookmark[0]{Supplementary Information}{supplementary-information}
\begin{center}
{\LARGE\bfseries Supplementary Information\par}
\vspace{0.5em}
{\Large\bfseries High-fidelity control of long-lived data qubits via a noisy mediator\par}
\vspace{1em}
{\large Nguyen H. Le, Tikai Chang, Eran Ginossar and Michael Stern\par}
\end{center}
\vspace{1em}

\section*{Supplementary Note 1: Cross-resonance implementation}
This section presents the explicit drive protocol used to realize the effective mediator-data interactions required for the single- and two-qubit gates in the reinitialization method.

\subsection*{Effective Hamiltonian and Lindbladian}

An effective $ZX$ interaction can be generated using a cross-resonance drive. The mediator qubit has frequency $\omega_m$ and is driven at the resonance frequency $\omega$ of the data qubit. The drive Hamiltonian is
\begin{equation}
H_d(t)={\cal E}X_m\cos\omega t,
\end{equation}
where ${\cal E}$ is the drive amplitude. The mediator and data qubit are coupled through a native $XX$ interaction of strength $g$. The total Hamiltonian is
\begin{equation}
H(t)=-\frac{\omega_m}{2}Z_m-\frac{\omega}{2}Z+gX_mX+H_d(t),
\end{equation}
where the first two terms are the bare mediator and data-qubit Hamiltonians.

A major noise source in many qubit platforms is $1/f$ noise, which produces fluctuations in the mediator frequency $\omega_m$. This noise can be modeled by
\begin{equation}
H_\chi(t)=\chi(t)Z_m,
\end{equation}
where $\chi(t)$ is a stochastic frequency-fluctuation signal. The mediator also undergoes relaxation at rate $\gamma_1=1/T_1$. Since the data qubits are assumed to have coherence times much longer than the gate duration, their intrinsic decoherence is neglected. The joint density matrix obeys
\begin{equation}
\frac{d\rho}{dt}=-i[H+H_\chi,\rho]+\mathcal L[\rho],
\end{equation}
with
\begin{equation}
\mathcal L[\rho]
=\gamma_1\left(
L_m\rho L_m^\dagger
-\frac12\left\{L_m^\dagger L_m,\rho\right\}
\right),
\end{equation}
where
\begin{equation}
L_m=\frac{X_m+iY_m}{2}\equiv\ket0_m\!\bra1_m
\end{equation}
is the mediator relaxation operator.

Let $\Delta=\omega-\omega_m$ be the detuning between the data qubit and the mediator. Typical parameters, chosen for the hybrid system of impurity spin donor qubits mediated by flux qubits, satisfy
\begin{equation}
\omega,\omega_m\gg |\Delta|\sim{\cal E}\gg g\sim \gamma_1.
\end{equation}
For example, in spin qubits mediated by flux qubits, the qubit frequencies are in the GHz range, the detunings and drive amplitudes are in the hundreds of MHz range, and $g$ is in the hundreds of kHz range. In this regime, the mediator relaxation rate $\gamma_1$ can be comparable to $g$.

We first move to the rotating frame defined by
\begin{equation}
U_0(t)=\exp\left\{i\frac{\omega t}{2}(Z_m+Z)\right\}.
\end{equation}
Within the rotating-wave approximation, where terms oscillating at $2\omega$ are dropped, the Hamiltonian becomes
\begin{equation}\label{eq:H1}
H_1=\frac{\Delta}{2}Z_m+\frac{{\cal E}}{2}X_m+\frac g2(X_mX+Y_mY)+\chi(t)Z_m.
\end{equation}
The dissipator is unchanged by this transformation,
\begin{equation}\label{eq:L1}
\mathcal L_1[\rho]=\mathcal L[\rho].
\end{equation}

Next introduce a tilted mediator basis
\begin{align}
Z_m'&=Z_m\cos\eta+X_m\sin\eta,\\
X_m'&=X_m\cos\eta-Z_m\sin\eta,\\
Y_m'&=Y_m,
\end{align}
where
\begin{equation}
\cos\eta=\frac{\Delta}{\sqrt{\Delta^2+{\cal E}^2}},
\qquad
\sin\eta=\frac{{\cal E}}{\sqrt{\Delta^2+{\cal E}^2}},
\end{equation}
which is obtained from the original basis by a rotation through an angle $\eta$ around the $Y_m$ axis.
In this basis,
\begin{align}
H_1=&
\frac{\Delta'}{2}Z_m'
+\frac g2\left(\cos\eta\,X_m'X+Y_m'Y\right)
+\frac g2\sin\eta\,Z_m'X\nonumber \\
&+\chi(t)(Z_m'\cos\eta-X_m'\sin\eta),
\end{align}
where
\begin{equation}
\Delta'=\sqrt{\Delta^2+{\cal E}^2}
\end{equation}
is the effective mediator splitting in the tilted frame.

Moving to the doubly rotating frame generated by
\begin{equation}
U_1(t)=\exp\left\{i\frac{\Delta't}{2}Z_m'\right\},
\end{equation}
the Hamiltonian becomes
\begin{equation}
H_2=\frac g2\sin\eta\,Z_m'X+\chi(t)\cos\eta\,Z_m'+V_{\rm cr}(t).
\end{equation}
This contains the desired effective $ZX$ interaction, a partially suppressed dephasing-noise term, and counter-rotating terms
\begin{align}
V_{\rm cr}(t)
&=\frac g2
\Big[
\sin\eta\left(X_m'\cos\Delta't-Y_m'\sin\Delta't\right)X
\nonumber\\
&\qquad\qquad+
\left(Y_m'\cos\Delta't+X_m'\sin\Delta't\right)Y
\Big]
\nonumber\\
&\quad-
\chi(t)\sin\eta\left(X_m'\cos\Delta't-Y_m'\sin\Delta't\right),
\end{align}
where the $1/f$ noise $\chi(t)$ is dominated by low-frequency components and is thus slowly varying on the scale of $2\pi/\Delta'$.

In the tilted and doubly rotating frame, the dissipator becomes
\begin{equation}
\mathcal L_2[\rho]
=
\sum_{\mu=x,y,z,L}\gamma_\mu
\left(
\Lambda_\mu'\rho \Lambda_\mu'^\dagger
-
\frac12\left\{\Lambda_\mu'^\dagger \Lambda_\mu',\rho\right\}
\right)
+
\mathcal L_{\rm cr}(t)[\rho],
\end{equation}
where the jump operators and rates are
\begin{align}
X_m' &: \gamma_x=\frac{\gamma_1}{8}(1-\cos\eta)^2,\\
Y_m' &: \gamma_y=\frac{\gamma_1}{8}(1-\cos\eta)^2,\\
Z_m' &: \gamma_z=\frac{\gamma_1}{4}\sin^2\eta,\\
L_m' &: \gamma_L=\gamma_1\cos\eta,
\end{align}
with $L_m'=(X_m'+iY_m')/2$. The counter-rotating part of the dissipator is
\begin{align}
\mathcal L_{\rm cr}(t)[\rho]
=&
\frac{\gamma_1\sin\eta}{4}
\bigg[
\sin\eta\,e^{2i\Delta't}L_m'\rho L_m'
\nonumber \\
&+(\cos\eta+1)e^{i\Delta't}L_m'\rho Z_m'
+{\rm h.c.}
\bigg].
\end{align}

The noise term $\chi(t)\cos\eta\,Z_m'$ can be modeled as pure dephasing in the tilted frame. Its effective rate is
\begin{equation}
\gamma_2\cos\eta=\frac{\cos\eta}{2T_\phi},
\end{equation}
where $\gamma_2=1/(2T_\phi)$ is the pure-dephasing rate in the absence of the drive. Thus, the total $Z_m'$ jump rate, including both relaxation induced and frequency-fluctuation-induced dephasing, is
\[
\gamma_z=\frac{\gamma_1}{4}\sin^2\eta+\gamma_2\cos\eta.
\]

Under the rotating-wave approximation, $V_{\rm cr}(t)$ and $\mathcal L_{\rm cr}(t)$ are neglected. Note that the $1/f$ noise $\chi(t)$ is slowly varying; hence, the last term in $V_{\rm cr}(t)$ is also rapidly oscillating and  negligible. The system is then described by the effective Hamiltonian
\begin{equation}
\tilde H=\tilde g Z_m'X,
\end{equation}
where
\begin{equation}
\tilde g=\frac g2\sin\eta=\frac g2\frac{{\cal E}}{\sqrt{\Delta^2+{\cal E}^2}},
\end{equation}
and the effective dissipator
\begin{equation}
\tilde{\mathcal L}[\rho]
=
\sum_{\mu=x,y,z,L}\gamma_\mu
\left(
\Lambda_\mu'\rho \Lambda_\mu'^\dagger
-
\frac12\left\{\Lambda_\mu'^\dagger \Lambda_\mu',\rho\right\}
\right),
\end{equation}
with jump operators and rates
\begin{align}
X_m' &: \gamma_x=\frac{\gamma_1}{8}(1-\cos\eta)^2,\\
Y_m' &: \gamma_y=\frac{\gamma_1}{8}(1-\cos\eta)^2,\\
Z_m' &: \gamma_z=\frac{\gamma_1}{4}\sin^2\eta+\gamma_2\cos\eta,\\
L_m' &: \gamma_L=\gamma_1\cos\eta.
\end{align}
This is precisely the general error channel considered in the main text. Therefore, the trajectory analysis and analytic fidelity formulas derived there apply directly to gates implemented by the cross-resonance drive.
\subsection*{Suppression of counter-rotating error}

The analytic estimate and  numerical simulation of the single-qubit gate infidelity in Fig.~3 of the main text use the rotating-wave approximation, where the counter-rotating terms $V_{\rm cr}(t)$ and $\mathcal L_{\rm cr}(t)$ are neglected. As these terms oscillate with frequency $\Delta'$ and $2\Delta'$, they can be suppressed to first order when the duration of the drive in each segment satisfies the commensurate condition
\begin{equation}
\tau = q \frac{2\pi}{\Delta'},
\end{equation}
where $q$ is a positive integer. At these durations, the oscillations cancel  over complete cycles.

For the single-qubit gate, substituting
\begin{equation}
\tau=\frac{\varphi_T}{M\tilde{g}(1-2\zeta)},
\end{equation}
where $\varphi_T$ is the target rotation phase, and
\begin{equation}
\tilde{g}=\frac{g}{2}\frac{{\cal E}}{\Delta'}, \quad
\Delta'=\sqrt{\Delta^2+{\cal E}^2},
\end{equation}
one obtains the commensurate condition
\begin{equation}
\Delta^2+{\cal E}^2=2qCg{\cal E},
\end{equation}
where
\begin{equation}
C=\frac{\pi M(1-2\zeta)}{2\varphi_T}.
\end{equation}
Solving for drive amplitude yields
\begin{equation}\label{eq:commensurate}
{\cal E}=q C g \pm \sqrt{(q C g)^2-\Delta^2}.
\end{equation}
In an experiment, the amplitude can  be swept until the commensurate condition is met.

For the two-qubit gate, assuming $\tilde{g}_1=\tilde{g}_2=\tilde{g}$, the duration of each segment is given by (see the main text)
\begin{equation}
\tau=\frac{\theta}{\tilde{g}}=\frac{1}{\tilde{g}}\sqrt{\frac{\varphi_T}{2M(1-2\zeta)}},
\end{equation}
and similar algebra shows that the commensurate condition is again given by equation~\eqref{eq:commensurate}, but with
\begin{equation}
C=\pi\sqrt{\frac{M (1-2\zeta)}{2\varphi_T}}.
\end{equation}

Numerical simulations with the drive amplitude adjusted to satisfy the commensurate condition above show that the counter-rotating terms do not lead to any noticeable change in the result of Fig.~3 of the main text.

\section*{Supplementary Note 2: Initialization into a non-diagonal state}
In the single- and two-qubit analyses of the main text, it is assumed that the mediator is initialized in a state whose density matrix is diagonal in the relevant basis. We now show that small off-diagonal terms do not affect the leading-order data-qubit dynamics.

First consider the single-qubit gate. The effective Hamiltonian between the mediator and a data qubit is
\begin{equation}
\tilde H=\tilde g Z_mX.
\end{equation}
Let the mediator be initialized in $\ket0_m$ with error probability $\zeta$, and allow small coherences of magnitude $|\xi|\ll1$,
\begin{align}\label{eq:rhoinit}
\rho_m^{\rm init}
=&(1-\zeta)\ket0_m\!\bra0
+\zeta\ket1_m\!\bra1
\nonumber \\
&+\xi\ket0_m\!\bra1
+\xi^*\ket1_m\!\bra0.
\end{align}
This is the most general single-qubit density matrix in this basis. If the data qubit is initially in the state $\rho_d(0)$, then
\begin{equation}
\rho(0)=\rho_m^{\rm init}\otimes\rho_d(0).
\end{equation}

Consider first trajectories with no mediator quantum jump. After a segment of duration $\tau$, the joint state is
\begin{align}\label{eq:rhof}
\rho(\tau)&=e^{-i\tilde H\tau}\rho(0)e^{i\tilde H\tau}\nonumber\\
&=(1-\zeta)\ket0_m\!\bra0\otimes e^{-i\theta X}\rho_d(0)e^{i\theta X}
\nonumber \\
&\quad+\zeta\ket1_m\!\bra1\otimes e^{i\theta X}\rho_d(0)e^{-i\theta X}\nonumber\\
&\quad+
\left[
\xi\ket0_m\!\bra1\otimes e^{-i\theta X}\rho_d(0)e^{-i\theta X}+{\rm h.c.}
\right],
\end{align}
where
\begin{equation}
\theta=\tilde g\tau.
\end{equation}
Tracing out the mediator gives
\begin{equation}\label{eq:rhodf}
\rho_d(\tau)=(1-\zeta)e^{-i\theta X}\rho_d(0)e^{i\theta X}
+\zeta e^{i\theta X}\rho_d(0)e^{-i\theta X}.
\end{equation}
The off-diagonal terms in the mediator density matrix do not contribute because their partial trace over the mediator vanishes. Equation~\eqref{eq:rhodf} can again be interpreted as a mixture of two trajectories: the intended rotation $e^{-i\theta X}$ with probability $1-\zeta$, and the opposite rotation $e^{i\theta X}$ with probability $\zeta$.

For trajectories containing a mediator jump, the off-diagonal terms contribute only through higher-order processes with probabilities of order $\mathcal O(\xi\gamma_\mu\tau_k)$. These terms are negligible at leading order. The same reasoning applies to the two-qubit gate when the mediator initial state is
\begin{align}
\rho_m^{\rm init}
=&(1-\zeta)\ket+_m\!\bra+ \
+\ \zeta\ket-_m\!\bra-
\nonumber \\
&+\xi\ket+_m\!\bra-
\ +\ \xi^*\ket-_m\!\bra+.
\end{align}
Thus small off-diagonal components in the mediator initial state do not change the leading-order trajectory analysis.

\section*{Supplementary Note 3: Trajectories and fidelity of the two-qubit gate}
We consider a mediator qubit coupled to two data qubits through the tunable effective Hamiltonians
\[
\tilde{H}_{1}=\tilde{g}_{1} Z_m X_1,
\qquad
\tilde{H}_{2}=\tilde{g}_{2} Z_m X_2,
\]
and assume single-qubit rotations of the mediator qubit generated by $X_m$ and $Y_m$ are also available. As described in the main text, each segment of the two-qubit gate is implemented using the four-pulse sequence
\begin{equation}
W=e^{-i\theta Y_m X_1}e^{-i\theta Z_m X_2}e^{i\theta Y_m X_1}e^{i\theta Z_m X_2}.
\end{equation}
From the second-order Baker-Campbell-Hausdorff formula
\begin{equation}
e^{A}e^{B}\approx e^{A+B+[A,B]/2},
\end{equation}
it follows that
\begin{equation}
W\approx e^{-2i\theta^2X_mX_1X_2}+\mathcal O(\theta^3).
\end{equation}
If the mediator is initialized in $\ket{+}_m$, this induces the data-qubit rotation
\begin{equation}
U=e^{-2i\theta^2X_1X_2}.
\end{equation}
This is the ideal case with no mediator decoherence.

The $Y_mX_1$ rotation can be implemented by conjugating the available $Z_mX_1$ evolution with single-qubit $\pm\pi/2$ rotations of the mediator about the $X_m$ axis:
\begin{equation}
 e^{-i\theta Y_m X_1}=R_m^\dagger e^{-i\theta Z_m X_1}R_m,
\end{equation}
where
\begin{equation}
R_m=e^{-i(\pi/4)X_m}.
\end{equation}
Similarly, the inverse evolutions can be implemented as
\begin{equation}
 e^{i\theta Y_mX_1}=R_m e^{-i\theta Z_m X_1}R_m^\dagger,
\end{equation}
and
\begin{equation}
 e^{i\theta Z_mX_2}=X_m e^{-i\theta Z_m X_2}X_m.
 \end{equation}
Thus, the full sequence $W$ can be written in terms of the available $Z_mX_k$ evolutions and mediator rotations as
\begin{align}\label{eq:W}
&R_m^\dagger e^{-i\theta Z_mX_1}R_m e^{-i\theta Z_m X_2} R_m e^{-i\theta Z_m X_1}R_m^\dagger \nonumber \\
&\times X_m e^{-i\theta Z_m X_2}X_m .
\end{align}
The two-qubit evolutions $e^{-i\theta Z_mX_1}$ and $e^{-i\theta Z_mX_2}$ are generated by activating $\tilde H_1$ and $\tilde H_2$, respectively, for the durations
\begin{equation}
\tau_1=\frac{\theta}{\tilde g_1},
\qquad
\tau_2=\frac{\theta}{\tilde g_2}.
\end{equation}

We now analyze the data-qubit trajectories in the presence of mediator initialization errors and mediator quantum jumps. The mediator is initialized in the intended state $\ket{+}_m$ with probability $1-\zeta$ and in the erroneous state $\ket{-}_m$ with probability $\zeta$, so that
\begin{equation}
\rho_m^{\rm init}=(1-\zeta)\ket{+}_m\!\bra{+}+\zeta\ket{-}_m\!\bra{-}.
\end{equation}

Mediator decoherence is described by jump operators $A_\mu\in\{X_m,Y_m,Z_m,L_m\}$ with rates $\gamma_x, \gamma_y,\gamma_z,\gamma_L$, respectively. These jumps may occur during any of the four $Z_mX_k$ evolution periods in the sequence. We assume $\zeta\ll 1$ and $\gamma_\mu\tau_k\ll1$, so trajectories containing more than one error event are higher order and can be neglected.

Let the data qubits initially be in an arbitrary state $\ket\psi$. If an initialization error occurs and  the mediator is prepared in $\ket{-}_m$, the state after the pulse sequence is then
\begin{equation}
W\ket{-}_m\otimes\ket\psi
\approx
\ket{-}_m\otimes e^{2i\theta^2X_1X_2}\ket\psi.
\end{equation}
Thus the data qubits undergo the same two-qubit rotation as in the ideal trajectory, but with the opposite sign.

Next consider a mediator jump during one of the $Z_mX_k$ pulses. Single-qubit rotations on the mediator are assumed to be much faster than the two-qubit $Z_mX_k$ pulses, so mediator decoherence during these single-qubit rotations is neglected. As a first example, suppose that an $X_m$ jump occurs within the interval $[t,t+dt]$, $0\le t\le\tau_1$, during the first (from left to right) $Z_mX_1$ pulse with probability $\gamma_x dt$, the interrupted evolution is
\begin{align}
 & e^{-i\tilde g_1(\tau_1-t)Z_mX_1}X_m e^{-i\tilde g_1tZ_mX_1}\nonumber \\&=X_m e^{i\tilde g_1(\tau_1-t)Z_mX_1}e^{-i\tilde g_1tZ_mX_1}\nonumber \\
 &=X_m e^{2i\theta(1-t/\tau_1)Z_mX_1}e^{-i\theta Z_mX_1}.
\end{align}
Thus the $X_m$ jump produces the additional error factor
\begin{equation}
X_m e^{2i\theta(1-t/\tau_1)Z_mX_1}.
\end{equation}

Using
\begin{equation}
R_m^\dagger X_m e^{2i\theta(1-t/\tau_1)Z_mX_1}
=X_m e^{2i\theta(1-t/\tau_1)Y_mX_1}R_m^\dagger,
\end{equation}
this error factor can be propagated to the left of the full pulse sequence. The corresponding Kraus operator is
\begin{align}
K_{X_m}^{(1,t)}
&=X_m e^{2i\theta(1-t/\tau_1)Y_mX_1}W\nonumber \\
&=X_m e^{2i\theta(1-t/\tau_1)Y_mX_1}e^{-2i\theta^2X_mX_1X_2}+\mathcal O(\theta^3).
\end{align}
Acting on $\ket{+}_m\otimes\ket\psi$, the final state at the end of a gate segment is
\begin{align}
&K_{X_m}^{(1,t)}\ket{+}_m\otimes\ket\psi\nonumber \\
&=
X_m e^{2i\theta(1-t/\tau_1)Y_mX_1}\ket{+}_m
\otimes e^{-2i\theta^2X_1X_2}\ket\psi.
\end{align}
Expanding the mediator state in the $\pm1$ eigenbasis of $Y_m$,
\begin{equation}
\ket{+}_m=\frac{1-i}{2}\ket{+y}_m+\frac{1+i}{2}\ket{-y}_m,
\end{equation}
the final state becomes
\begin{align}
&\frac{1-i}{2}X_m\ket{+y}_m e^{2i\theta(1-t/\tau_1)X_1}e^{-2i\theta^2X_1X_2}\ket\psi+\nonumber \\
&
\frac{1+i}{2}X_m\ket{-y}_m e^{-2i\theta(1-t/\tau_1)X_1}e^{-2i\theta^2X_1X_2}\ket\psi.
\end{align}
After tracing out the mediator, the reduced data-qubit state is
\begin{equation}
\rho_d=\frac12 U_+\ket\psi\!\bra\psi U_+^\dagger+\frac12 U_-\ket\psi\!\bra\psi U_-^\dagger,
\end{equation}
where
\begin{equation}
U_\pm=e^{\pm2i\theta(1-t/\tau_1)X_1}e^{-2i\theta^2X_1X_2}.
\end{equation}
The data-qubit trajectory is therefore a mixture of two unitary branches, each occurring with probability $\gamma_xdt/2$. In addition to the intended two-qubit rotation, the jump produces a coherent single-qubit error on the first data qubit.

The same procedure can be applied to $X_m$, $Y_m$, $Z_m$, and $L_m$ jumps during the second, third, and fourth (from left to right) $Z_mX_k$ pulses. Using the Baker--Campbell--Hausdorff formula to second order in $\theta$ to propagate the error term out of the full sequence, the Kraus operators for all cases are
\begin{align}
K_{X_m}^{(1, t)}&=X_me^{2i\theta (1-t/\tau_1)Y_mX_1}e^{-2i\theta^2 X_m X_1X_2},\nonumber \\
K_{X_m}^{(2, t)}&=X_m e^{2i\theta Y_mX_1}
e^{2i\theta(1-t/\tau_2)Z_mX_2}
e^{2i\theta^2(1-2t/\tau_2)X_mX_1X_2},\nonumber \\
K_{X_m}^{(3, t)}&=X_m e^{2i\theta(t/\tau_1)Y_mX_1}
e^{2i\theta Z_mX_2}
e^{2i\theta^2(2t/\tau_1-1)X_mX_1X_2},\nonumber \\
K_{X_m}^{(4, t)}&=X_m e^{2i\theta(t/\tau_2) Z_mX_2}
e^{-2i\theta^2X_mX_1X_2},\nonumber \\
K_{Y_m}^{(1, t)}&=-Z_me^{2i\theta(1-t/\tau_1)Y_mX_1}
e^{-2i\theta^2X_mX_1X_2},\nonumber \\
K_{Y_m}^{(2, t)}&=Y_m e^{2i\theta(1-t/\tau_2)Z_mX_2}
e^{2i\theta^2(1-2t/\tau_2)X_mX_1X_2},\nonumber \\
K_{Y_m}^{(3, t)}&=Z_m e^{2i\theta(t/\tau_1)Y_mX_1}
e^{2i\theta^2(1-2t/\tau_1)X_mX_1X_2},\nonumber \\
K_{Y_m}^{(4, t)}&=-Y_m e^{2i\theta(t/\tau_2) Z_mX_2}
e^{2i\theta^2X_mX_1X_2},\nonumber \\
K_{Z_m}^{(1, t)}&=Y_m e^{-2i\theta^2 X_mX_1X_2},\nonumber \\
K_{Z_m}^{(2, t)}&= Z_m e^{2i\theta Y_mX_1}e^{-2i\theta^2 X_mX_1X_2},\nonumber \\
K_{Z_m}^{(3, t)}&= -Y_m e^{2i\theta Z_m X_2}e^{2i\theta^2 X_mX_1X_2} ,\nonumber \\
K_{Z_m}^{(4, t)}&=-Z_m e^{2i\theta^2 X_mX_1X_2},\nonumber \\
K_{L_m}^{(1, t)}&=\frac12K_{X_m}^{(1, t)}+\frac{i}{2}K_{Y_m}^{(1, t)},\nonumber \\
K_{L_m}^{(2, t)}&=\frac12K_{X_m}^{(2, t)}+\frac{i}{2}K_{Y_m}^{(2,t)},\nonumber \\
K_{L_m}^{(3, t)}&=\frac12K_{X_m}^{(3, t)}+\frac{i}{2}K_{Y_m}^{(3, t)},\nonumber \\
K_{L_m}^{(4, t)}&=\frac12K_{X_m}^{(4, t)}+\frac{i}{2}K_{Y_m}^{(4, t)},
\end{align}
where the last four Kraus operators for the $L_m$ error follow from $L_m=(X_m+iY_m)/2$.
These Kraus operators are unitary except  the last four cases. Note that here ``first'', ``second'', ``third'', and ``fourth'' refer to the left-to-right order in the pulse sequence, rather than the order of occurrence in time.

Applying these operators to the initial state of the total system and tracing out the mediator gives a reduced data-qubit state that can be interpreted, to leading order, as a mixture of  trajectory branches, each characterized by a unitary operator
\begin{equation}
U_{q,r,s}=e^{2iq\theta X_1}e^{2ir\theta X_2}e^{-2is\theta^2X_1X_2}+\mathcal O(\theta^3),
\end{equation}
where $q$, $r$, and $s$ are random parameters determined by the type, timing, and location of the error event. The full list of first-order trajectories and their probabilities is shown in Table~\ref{tab:traj}. Mediator initialization and decoherence therefore produce two effects on the data qubits: coherent single-qubit errors generated by $X_1$ and $X_2$, and fluctuations in the two-qubit rotation angle.
\begin{table}[p]
    \centering
    \scriptsize
    \setlength{\tabcolsep}{3pt}
\begin{tabular}{|c|c|c|c|c|}
\hline
 $q$ &$r$ &$s$ & probability & error type \\ \hline
$0$ &$0$ &$-1$ & $\zeta$ &  initialization error \\ \hline
$1-t/\tau_1$ &$0$ &$1$ &$\gamma_x dt/2$, $(0\leq t\leq \tau_1)$ & $X$ in 1st pulse, 1st branch \\ \hline
$t/\tau_1-1$&$0$ &$1$ &$\gamma_x dt/2$, $(0\leq t\leq \tau_1)$ & $X$ in 1st pulse, 2nd branch \\ \hline
$1$&$t/\tau_2-1$ & $2t/\tau_2-1$& $\gamma_x dt/4$, $(0\leq t\leq \tau_2)$& $X$ in 2nd pulse, 1st branch \\ \hline
$1$&$1-t/\tau_2$ & $2t/\tau_2-1$& $\gamma_x dt/4$, $(0\leq t\leq \tau_2)$ & $X$ in 2nd pulse, 2nd branch \\ \hline
$-1$&$t/\tau_2-1$ & $2t/\tau_2-1$& $\gamma_x dt/4$, $(0\leq t\leq \tau_2)$& $X$ in 2nd pulse, 3rd branch \\ \hline
$-1$&$1-t/\tau_2$ & $2t/\tau_2-1$&$\gamma_x dt/4$, $(0\leq t\leq \tau_2)$ & $X$ in 2nd pulse, 4th branch \\ \hline
$t/\tau_1$& $1$&$1-2t/\tau_1$ &$\gamma_x dt/4$, $(0\leq t\leq \tau_1)$ &$X$ in 3rd pulse, 1st branch\\ \hline
$t/\tau_1$& $-1$ &$1-2t/\tau_1$ & $\gamma_x dt/4$, $(0\leq t\leq \tau_1)$& $X$ in 3rd pulse, 2nd branch \\ \hline
$-t/\tau_1$& $1$&$1-2t/\tau_1$ &$\gamma_x dt/4$, $(0\leq t\leq \tau_1)$ &$X$ in 3rd pulse, 3rd branch\\ \hline
$-t/\tau_1$& $-1$ &$1-2t/\tau_1$ & $\gamma_x dt/4$, $(0\leq t\leq \tau_1)$& $X$ in 3rd pulse, 4th branch \\ \hline
$0$ &$t/\tau_2$ &$1$ &$\gamma_x dt/2$, $(0\leq t\leq \tau_2)$  & $X$ in 4th pulse, 1st branch\\ \hline
$0$& $-t/\tau_2$&$1$ & $\gamma_x dt/2$, $(0\leq t\leq \tau_2)$& $X$ in 4th pulse, 2nd branch\\ \hline
$t/\tau_1-1$ &$0$ &$1$ &$\gamma_y dt/2$, $(0\leq t\leq \tau_1)$ & $Y$ in 1st pulse, 1st branch \\ \hline
$1-t/\tau_1$&$0$ &$1$ &$\gamma_y dt/2$, $(0\leq t\leq \tau_1)$ & $Y$ in 1st pulse, 2nd branch \\ \hline
$0$&$t/\tau_2-1$ & $2t/\tau_2-1$& $\gamma_y dt/2$, $(0\leq t\leq \tau_2)$& $Y$ in 2nd pulse, 1st branch \\ \hline
$0$&$1-t/\tau_2$ & $2t/\tau_2-1$& $\gamma_y dt/2$, $(0\leq t\leq \tau_2)$ & $Y$ in 2nd pulse, 2nd branch \\ \hline
$t/\tau_1$& $0$&$2t/\tau_1-1$ &$\gamma_y dt/2$, $(0\leq t\leq \tau_1)$ &$Y$ in 3rd pulse, 1st branch \\ \hline
$-t/\tau_1$& $0$ &$2t/\tau_1-1$ & $\gamma_y dt/2$, $(0\leq t\leq \tau_1)$& $Y$ in 3rd pulse, 2nd branch \\ \hline
$0$ &$t/\tau_2$ &$-1$ &$\gamma_y dt/2$, $(0\leq t\leq \tau_2)$  & $Y$ in 4th pulse, 1st branch\\ \hline
$0$& $-t/\tau_2$&$-1$ & $\gamma_y dt/2$, $(0\leq t\leq \tau_2)$& $Y$ in 4th pulse, 2nd branch\\ \hline
$0$ &$0$ & $1$ &$\gamma_z \tau_1$ & $Z$ in 1st pulse\\ \hline
$1$& $0$& $1$& $\gamma_z \tau_2/2$& $Z$ in 2nd pulse, 1st branch\\ \hline
$-1$&$0$ &$1$ &$\gamma_z \tau_2/2$  & $Z$ in 2nd pulse, 2nd branch\\ \hline
$0$ &$1$ &$-1$ &$\gamma_z \tau_1/2$ &$Z$ in 3rd pulse, 1st branch \\ \hline
$0$& $-1$ &$-1$ & $\gamma_z \tau_1/2$ & $Z$ in 3rd pulse, 2nd branch\\ \hline
$0$ &$0$ &$-1$ &$\gamma_z \tau_2$ & $Z$ in 4th pulse\\ \hline
$t/\tau_1-1$&$0$ &$1$ &$\gamma_L dt/2$, $(0\leq t\leq \tau_1)$ & $L$ in 1st pulse \\ \hline
$0$& $1-t/\tau_2$ &$2t/\tau_2-1$ & $\gamma_L dt/4$, $(0\leq t\leq \tau_2)$& $L$ in 2nd pulse, 1st branch \\ \hline
$1$& $1-t/\tau_2$ &$2t/\tau_2-1$ & $\gamma_L dt/8$, $(0\leq t\leq \tau_2)$& $L$ in 2nd pulse, 2nd branch \\ \hline
$-1$& $1-t/\tau_2$ &$2t/\tau_2-1$ & $\gamma_L dt/8$, $(0\leq t\leq \tau_2)$& $L$ in 2nd pulse, 3rd branch \\ \hline
$t/\tau_1$& $0$ &$2t/\tau_1-1$ & $\gamma_L dt/8$, $(0\leq t\leq \tau_1)$& $L$ in 3rd pulse, 1st branch \\ \hline
$t/\tau_1$& $0$ &$0$ & $\gamma_L dt/8$, $(0\leq t\leq \tau_1)$& $L$ in 3rd pulse, 2nd branch \\ \hline
$t/\tau_1$& $1$ &$1-2t/\tau_1$ & $\gamma_L dt/16$, $(0\leq t\leq \tau_1)$& $L$ in 3rd pulse, 3rd branch \\ \hline
$t/\tau_1$& $1$ &$0$ & $\gamma_L dt/16$, $(0\leq t\leq \tau_1)$& $L$ in 3rd pulse, 4th branch \\ \hline
$t/\tau_1$& $-1$ &$1-2t/\tau_1$ & $\gamma_L dt/16$, $(0\leq t\leq \tau_1)$& $L$ in 3rd pulse, 5th branch \\ \hline
$t/\tau_1$& $-1$ &$0$ & $\gamma_L dt/16$, $(0\leq t\leq \tau_1)$& $L$ in 3rd pulse, 6th branch \\ \hline
$0$& $-t/\tau_2$&$1$ &$\gamma_L dt/2$, $(0\leq t\leq \tau_2)$ &$L$ in 4th pulse\\ \hline
$0$ &$0$ &$1$ & $1-\zeta-2\left(\gamma_x+\gamma_y+\gamma_z+\gamma_L/2\right)(\tau_1+\tau_2)$ & none \\ \hline
\end{tabular}
\caption{Complete list of data-qubit trajectories and probabilities for all first-order error processes during one segment of the two-qubit gate.}\label{tab:traj}
\end{table}

From Table~\ref{tab:traj}, the mean values of the trajectory parameters are
\begin{align}
\langle q\rangle&=\langle r\rangle=0,\nonumber\\
\langle s\rangle&=1-2\zeta
-\gamma_x(\tau_1+\tau_2)
-\gamma_y(\tau_1+3\tau_2)
-2\gamma_z(\tau_1+\tau_2)\nonumber \\
&\quad -\frac{\gamma_L}{2}(\tau_1+\tau_2).
\end{align}
To first order in $\zeta$ and $\gamma_\mu\tau_k$, the variances are
\begin{align}\label{eq:var}
\langle\delta q^2\rangle&=\gamma_x\left(\frac{2\tau_1}{3}+\tau_2\right)
+\frac23\gamma_y\tau_1
+\gamma_z\tau_2
+\gamma_L\left(\frac{\tau_1}{3}+\frac{\tau_2}{4}\right),\nonumber\\
\langle\delta r^2\rangle&
=\gamma_x\left(\tau_1+\frac{2\tau_2}{3}\right)
+\frac23\gamma_y\tau_2
+\gamma_z\tau_1
+\gamma_L\left(\frac{\tau_1}{4}+\frac{\tau_2}{3}\right),\nonumber\\
\langle\delta s^2\rangle&\approx
4\zeta
+\frac43\gamma_x(\tau_1+\tau_2)
+\gamma_y\left(\frac43\tau_1+\frac{16}{3}\tau_2\right)\nonumber \\
&\quad+4\gamma_z(\tau_1+\tau_2)
+\gamma_L\left(\frac{7}{12}\tau_1+\frac23\tau_2\right),
\end{align}
where $\delta q=q-\langle q\rangle$, and similarly for $\delta r$ and $\delta s$.

For $M$ segments, a trajectory is specified by the set of random variables $\bm v=\{q_j,r_j,s_j\}$ for $1\leq j \leq M$. The corresponding data-qubit unitary is
\begin{align}
U_{\bm v}=e^{-i\alpha_{\bm v}X_1}e^{-i\beta_{\bm v}X_2}e^{-i\varphi_{\bm v}X_1X_2},
\end{align}
with
\begin{align}\label{eq:angle}
\alpha_{\bm v}=2\theta\sum_j q_j,
\qquad
\beta_{\bm v}=2\theta\sum_j r_j,
\qquad
\varphi_{\bm v}=2\theta^2\sum_j s_j.
\end{align}
To implement the target gate
\begin{equation}
U_T=e^{-i\varphi_T X_1X_2},
\end{equation}
we choose $\tau_1$ and $\tau_2$ so that the average two-qubit rotation angle equals the target angle,
\begin{equation}
\langle\varphi\rangle=2M\theta^2\langle s\rangle=\varphi_T.
\end{equation}
Thus
\begin{equation}
\theta=\sqrt{\frac{\tilde\varphi_T}{2M}},
\end{equation}
where the renormalized target phase is
\begin{align}
\tilde\varphi_T=\frac{\varphi_T}{\langle s\rangle}
=&\varphi_T\bigg[
1-2\zeta
-\gamma_x(\tau_1+\tau_2)
-\gamma_y(\tau_1+3\tau_2)\nonumber \\
&-2\gamma_z(\tau_1+\tau_2)
-\frac{\gamma_L}{2}(\tau_1+\tau_2)
\bigg]^{-1}.
\end{align}

In the large-$M$ limit, the small decoherence probability within each segment is dominated by the initialization error, giving
\begin{equation}
\tilde\varphi_T\approx \frac{\varphi_T}{1-2\zeta}.
\end{equation}
This is the limit considered in the main text.

The corresponding $Z_mX_k$ pulse durations are
\begin{equation}\label{eq:tau2}
\tau_1=\frac{\theta}{\tilde g_1}\approx\frac1{\tilde g_1}\sqrt{\frac{\tilde\varphi_T}{2M}},
\qquad
\tau_2=\frac{\theta}{\tilde g_2}\approx\frac1{\tilde g_2}\sqrt{\frac{\tilde\varphi_T}{2M}}.
\end{equation}
Substituting into equation~\eqref{eq:angle} gives
\begin{align}
\alpha_{\bm v}&=\sqrt{\frac{2\tilde\varphi_T}{M}}\sum_{j=1}^M q_j,
\quad
\beta_{\bm v}=\sqrt{\frac{2\tilde\varphi_T}{M}}\sum_{j=1}^M r_j,
\nonumber \\
\varphi_{\bm v}&=\frac{\tilde\varphi_T}{M}\sum_{j=1}^M s_j.
\end{align}
Since $\alpha$ and $\beta$ have zero mean,
\begin{align}
\langle\delta\alpha^2\rangle&=\langle\alpha^2\rangle=2\tilde\varphi_T\langle\delta q^2\rangle,
\nonumber \\
\langle\delta\beta^2\rangle&=\langle\beta^2\rangle=2\tilde\varphi_T\langle\delta r^2\rangle,
\end{align}
and
\begin{align}
\langle\delta\varphi^2\rangle=\frac{\tilde\varphi_T^2}{M}\langle\delta s^2\rangle.
\end{align}

As shown for the single-qubit gate in the main text, the gate fidelity is the trajectory average
\begin{equation}
G=\mathbb E_{\bm v}[G_{\bm v}]=\sum_{\bm v}P(\bm v)G_{\bm v},
\end{equation}
where $P(\bm v)$ is the probability of trajectory $\bm v$, and $G_{\bm v}$ is the single-trajectory state fidelity averaged over Haar-random two-qubit input states
\begin{equation}
G_{\bm v}=\mathbb E_\psi[F_{\bm v}(\psi)].
\end{equation}
For a trajectory $\bm v$, the state fidelity is
\begin{align}
F_{\bm v}(\psi)&=\left|\bra\psi U_T^\dagger U_{\bm v}\ket\psi\right|^2\nonumber\\
&=\left|\bra\psi e^{-i\alpha_{\bm v}X_1}e^{-i\beta_{\bm v}X_2}e^{-i\delta\varphi_{\bm v}X_1X_2}\ket\psi\right|^2,
\end{align}
where $\delta\varphi_{\bm v}=\varphi_{\bm v}-\varphi_T=\varphi_{\bm v}-\langle\varphi\rangle$. Haar averaging gives, to second order in $\alpha_{\bm v}$, $\beta_{\bm v}$, and $\delta\varphi_{\bm v}$,
\begin{align}
G_{\bm v}\approx1-\frac45\left(\alpha_{\bm v}^2+\beta_{\bm v}^2+\delta\varphi_{\bm v}^2\right).
\end{align}
Averaging over trajectories gives
\begin{align}
G&=1-\frac45\left(\langle\alpha_{\bm v}^2\rangle+\langle\beta_{\bm v}^2\rangle+\langle\delta\varphi_{\bm v}^2\rangle\right)\nonumber \\
&=1-\frac45\left[2\tilde\varphi_T(\langle q^2\rangle+\langle r^2\rangle)+\frac{\tilde\varphi_T^2}{M}\langle\delta s^2\rangle\right].
\end{align}
Using equation~\eqref{eq:var}, the gate infidelity is
\begin{align}
1-G&\approx\frac85\tilde\varphi_T
\left(\frac53\gamma_x+\frac23\gamma_y+\gamma_z+\frac7{12}\gamma_L\right)(\tau_1+\tau_2)\nonumber\\
&+\frac45\frac{\tilde\varphi_T^2}{M}
\bigg[
4\zeta
+\frac43\gamma_x(\tau_1+\tau_2)
+\gamma_y\left(\frac43\tau_1+\frac{16}{3}\tau_2\right)\nonumber \\
& +4\gamma_z(\tau_1+\tau_2)
+\gamma_L\left(\frac7{12}\tau_1+\frac23\tau_2\right)
\bigg].
\end{align}
Substituting the pulse durations from equation~\eqref{eq:tau2}, this can be written as
\begin{align}
1-G=
&\frac{8\tilde\varphi_T^{3/2}}{5}\frac{\Gamma}{\bar g}\sqrt{\frac2M}
+
\frac{16\tilde\varphi_T^2}{5}\frac{\zeta}{M}\nonumber \\
&+
\frac{8\tilde\varphi_T^{5/2}}{5\sqrt2}\frac{\Gamma'}{\bar g}\frac1{M\sqrt M},
\end{align}
where
\begin{align}
\Gamma&=\frac53\gamma_x+\frac23\gamma_y+\gamma_z+\frac7{12}\gamma_L,\nonumber\\
\frac1{\bar g}&=\frac12\left(\frac1{g_1}+\frac1{g_2}\right),\nonumber\\
\Gamma'&=\frac43\gamma_x
+\gamma_y\left(\frac23\frac{\bar g}{g_1}+\frac83\frac{\bar g}{g_2}\right)
+4\gamma_z\nonumber \\
&\quad +\gamma_L\left(\frac7{24}\frac{\bar g}{g_1}+\frac13\frac{\bar g}{g_2}\right).
\end{align}
The $1/\sqrt M$ term arises from coherent single-qubit errors, represented by $\alpha$ and $\beta$, due to mediator decoherence. The $1/M$ term comes from fluctuations in the two-qubit rotation angle due to initialization error. The $1/(M\sqrt M)$ term comes from decoherence-induced fluctuations in the two-qubit angle. For large $M$, the last contribution is always dominated by the first, and is therefore neglected in the main text.

\section*{Supplementary Note 4: Fast mediator initialization via a resonator mode}
In systems where the mediator can be strongly coupled to a resonator, such as superconducting flux-qubit platforms, fast mediator initialization can be achieved by turning off the drive and tuning the mediator frequency close to a resonator mode. The strong flip-flop interaction between the mediator and resonator then transfers excitations from the mediator to the resonator.

The mediator-resonator Hamiltonian is
\begin{equation}
H_{mr}=-\frac{\omega_m}{2}Z_m+\omega_r a^\dagger a+\lambda X_m(a+a^\dagger),
\end{equation}
where $\lambda$ is the coupling strength. Changing the applied flux shifts the transverse energy of the flux qubit by $\epsilon$, giving the effective mediator frequency
\[
\omega_m'=\sqrt{\omega_m^2+\epsilon^2}.
\]
By sweeping this frequency toward the resonator frequency, the mediator-resonator coupling enables the exchange
\[
\ket{1,n}\leftrightarrow\ket{0,n+1},
\]
at a rate set by $\lambda$. In strongly coupled devices, values of $\lambda\sim2\pi\times100\,{\rm MHz}$ are achievable, giving excitation-transfer times on the order of $10\,{\rm ns}$.

Only a small excitation probability accumulates in the mediator during each short gate segment. Therefore, a full exchange cycle is not necessary: a shorter resonator-assisted reset pulse can be sufficient to reduce the mediator excitation probability back below the target threshold $\zeta$. For a lossy resonator mode with loss rate $\kappa\sim 2\pi\times 10$ MHz, the accumulated excitation is drained completely before the next reset. 

\section*{Supplementary Note 5: Effect of imperfections}
\subsection*{Resilience against parameter drift}

The proposed gate is resilient to several quasi-static parameter drifts. We model such drifts by the substitutions
\[
\omega_m\to\omega_m+d\omega_m,
\qquad
{\cal E}\to{\cal E}+d{\cal E},
\qquad
 g\to g+dg.
\]
In addition, a displacement of the mediator qubit's $Z$ axis, for example due to flux noise in a flux-based qubit, can be modeled by an additional transverse term $\epsilon X_m$ in the Hamiltonian. Here $d\omega_m$, $d{\cal E}$, $dg$, and $\epsilon$ denote quasi-static noise in the mediator frequency, drive amplitude, coupling strength, and transverse energy displacement, respectively.

For the cross-resonance protocol, the doubly rotating-frame analysis gives the following scaling estimates. First, the transverse perturbation $\epsilon X_m$ becomes a counter-rotating term with amplitude $\epsilon$ oscillating at the data-qubit frequency $\omega$. It therefore produces a unitary error of order $\mathcal O(\epsilon/\omega)$ and an infidelity of order $\mathcal O(\epsilon^2/\omega^2)$. Second, a fluctuation $d\omega_m$ in the mediator transition frequency produces terms with amplitude $d\omega_m$ oscillating at
\[
\Delta'=\sqrt{{\cal E}^2+\Delta^2},
\]
leading to an infidelity contribution of order $\mathcal O((d\omega_m)^2/\Delta'^2)$. Third, fluctuations in ${\cal E}$ and $g$ directly change the effective rotation rate
\[
\tilde g=\frac{g{\cal E}}{2\sqrt{{\cal E}^2+\Delta^2}}.
\]
The relative error in $\tilde g$ scales as
\[
\frac{d\tilde g}{\tilde g}
=
\left(\frac{\Delta}{\Delta'}\right)^2\frac{d{\cal E}}{{\cal E}}
+
\frac{dg}{g}.
\]
Thus,
\[
\frac{(d\tilde g)^2}{\tilde g^2}
\sim
\left(\frac{\Delta}{\Delta'}\right)^4\frac{(d{\cal E})^2}{{\cal E}^2}
+
\frac{(dg)^2}{g^2},
\]
up to a factor of at most 2. Combining these contributions, the gate infidelity from quasi-static noise scales as
\begin{align}\label{eq:dG}
dG\sim
&\mathcal O\!\left(\frac{\epsilon^2}{\omega^2}\right)
+
\mathcal O\!\left(\frac{(d\omega_m)^2}{\Delta'^2}\right)
+
\mathcal O\!\left(\frac{\Delta^4}{\Delta'^4}\frac{(d{\cal E})^2}{{\cal E}^2}\right)+
\mathcal O\!\left(\frac{(dg)^2}{g^2}\right).
\end{align}

For spin qubits mediated by flux qubits, transverse energy drifts are typically $\lesssim10\,{\rm MHz}$, while the data-qubit frequency is typically $\omega\sim10\,{\rm GHz}$. The first term in equation~\eqref{eq:dG} is therefore very small at $\sim 10^{-6}$. Because the mediator flux qubit is operated at an optimal flux point, fluctuations in its transition frequency are typically below 10 kHz; with detunings and drive amplitudes in the hundreds of MHz range, the second term is also negligible at $\sim 10^{-8}$.

  The error due to drift in the amplitude ${\cal E}$ is also greatly suppressed in the strong-drive regime where ${\cal E}\gg \Delta$ due to the $\Delta^4/\Delta'^4$ factor. For ${\cal E}/\Delta\sim 3$ and a typical drive-amplitude fluctuation of $1\%$, equation~\eqref{eq:dG} predicts a small  error $\sim10^{-6}$.  In addition, the coupling between the flux qubit and spin qubit arises from the interaction between the spin and the magnetic moment generated by the flux-qubit persistent current. Flux-noise-induced fluctuations in this current, and therefore in $g$, are typically of order $10^{-6}$ over a representative experiment. This gives an estimated infidelity contribution of order $10^{-12}$, which is also negligible.

\subsection*{Crosstalk with spectator qubits}

Multiple data qubits may share a common mediator within a sub-processor. We now analyze the effect of a spectator qubit on both the single- and two-qubit gates.

Consider a mediator coupled to a data qubit and a spectator qubit through
\begin{align}
H_{\rm ms}&= -\frac{\omega_m}{2}Z_m-\frac{\omega}{2}Z-\frac{\omega_s}{2}Z_s+gZ_mZ+g_sZ_mZ_s\nonumber \\
&\quad +{\cal E}X_m\cos\omega t,
\end{align}
where $\omega_s$ is the spectator frequency and $g_s$ is its coupling to the mediator. In the doubly rotating frame and within the RWA, the Hamiltonian becomes
\begin{equation}
\tilde H_{\rm ms}=\tilde g Z_m'X-\frac{\Delta_s}{2}Z_s+\tilde g_s Z_m'X_s,
\end{equation}
where $\Delta_s=\omega_s-\omega$ is the detuning between the spectator and the addressed data qubit.

When the mediator is prepared in an eigenstate of $Z_m'$, the mediator dynamics are effectively frozen. The data qubit and the spectator then evolve under separate effective Hamiltonians,
\begin{equation}
\tilde H_m=\tilde g X,
\qquad
\tilde H_s=-\frac{\Delta_s}{2}Z_s+\tilde g_sX_s.
\end{equation}
The data qubit therefore undergoes the intended $X$ rotation, with no backaction from the spectator.
\begin{figure}[t]
    \centering
\includegraphics[width=0.6\linewidth]{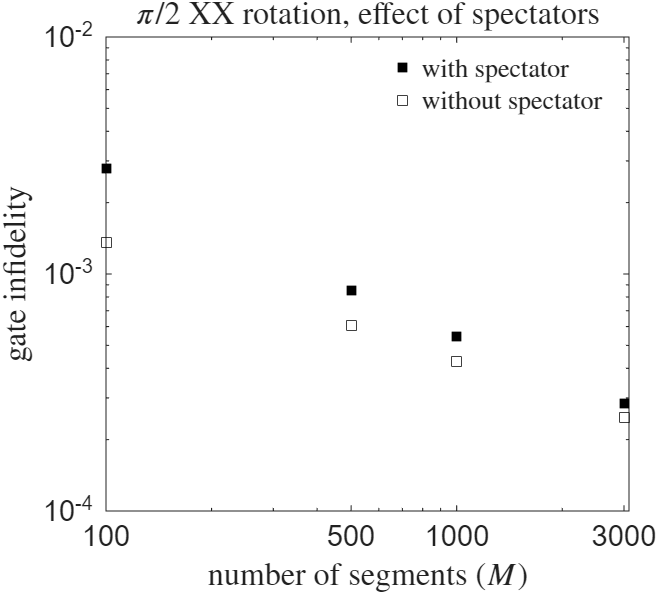}
\caption{\textbf{Influence of a spectrally close spectator qubit.} Two-qubit gate infidelity with and without a spectator qubit close in frequency to the second data qubit.}\label{fig:2qspec}
\end{figure}
For the spectator, moving to the interaction frame of the free term $-\Delta_sZ_s/2$ gives
\begin{equation}
\tilde H_s(t)=\tilde g_s\left(X_s\cos\Delta_s t+Y_s\sin\Delta_s t\right).
\end{equation}
If the spectator is spectrally well separated from the data qubit, $\Delta_s\gg\tilde g_s$, this term oscillates rapidly and averages to zero. If the spectator is spectrally close to the addressed data qubit, $\Delta_s\sim\tilde g_s$, the unwanted spectator rotation can either be avoided by qubit selection or tracked and compensated in later single-qubit gate cycles.

The spectator effect is more subtle for the two-qubit gate, because the spectator can participate in the pulse sequence and become entangled with the data qubits. However, this effect scales with the segment duration and is therefore expected to vanish in the large-$M$ limit. Figure~\ref{fig:2qspec} shows a numerical simulation of the two-qubit gate infidelity when a spectator is spectrally close to the second data qubit. The parameters are $\tilde g_s=\tilde g_2$ and $\omega_s-\omega_2=1.5\tilde g_s$, so the spectator participates during the $Z_mX_2$ pulse applied to the second data qubit. Other parameters are $\tilde{g}_1=\tilde{g}_2=2\pi \times 100 \ \rm kHz$,  and mediator relaxation time $T_1=100 \rm \mu s$; mediator dephasing time is neglected, assuming complete filtering by the strong drive. The difference between the infidelity with and without the spectator decreases as $M$ increases, confirming that the spectator has a negligible effect in the large-$M$ regime.

\subsection*{Supplementary Note 6: Parameter values for the cross-platform estimations}
\label{sec:supp-platform-parameters}

The estimates in Table~I in the main text are evaluated for
$M=100$ and $\varphi_T=\pi/4$. For unequal mediator--data-qubit couplings,
the effective coupling is defined as the harmonic mean,
\begin{equation}
    \frac{1}{\bar g}
    =
    \frac{1}{2}
    \left(
        \frac{1}{\widetilde g_1}
        +
        \frac{1}{\widetilde g_2}
    \right).
\end{equation}
The active gate duration, total duration, and mediator-relaxation
contribution to the gate infidelity are calculated using the analytic
infidelity formula derived for the two-qubit gate in the main text.

\subsubsection*{Nuclear-spin qubits mediated by a quantum-dot electron}

For the two ${}^{29}\mathrm{Si}$ nuclear spins, the reported hyperfine
couplings are $A_1/h=448.5~\mathrm{kHz}$, taken from the measurement of the 1st dot, and
$A_2/h=179.8~\mathrm{kHz}$~\cite{supp:Hensen2020}, taken from the measurement of the 2nd dot. Using
$\widetilde g_i/2\pi=|A_i|/(4h)$ gives the harmonic-mean coupling
$\bar g/2\pi=64.2~\mathrm{kHz}$. Because the electron relaxation time
was not reported for this device, we use the representative value
$T_{1}=0.88~\mathrm{ms}$ measured in a separate SiMOS quantum dot
near a spin--valley relaxation hotspot~\cite{supp:Zhang2020}. The reset time
$t_{\mathrm{reset}}=6.3~\mathrm{ns}$ is inferred from the demonstrated
spin--photon coupling $\lambda/2\pi\simeq40~\mathrm{MHz}$, corresponding to
the resonant half-swap time $\pi/(2\lambda)$~\cite{supp:HarveyCollard2022}. This
 is the duration required for a coherent transfer of the electron excitation to a
resonator.

For $M=100$ and target phase $\varphi_T=\pi/4$, the
active-interaction duration is
\begin{align}
    T_{\mathrm{active}}
    &=
    \frac{2\sqrt{2M\varphi_T}}{\bar g}
\simeq62.2~\mu\mathrm{s}.
\end{align}
Including the $M-1=99$ mediator resets gives
\begin{align}
    T_{\mathrm{tot}}
    &=
    T_{\mathrm{active}}
    +(M-1)t_{\mathrm{reset}}
    \simeq62.8~\mu\mathrm{s}.
\end{align}

For the relaxation-only estimate, we use
\begin{equation}
    \Gamma
    =
    \frac{13}{24T_{1}}
    \simeq
    6.16\times10^{2}~\mathrm{s}^{-1}.
\end{equation}
The residual mediator-relaxation contribution is then
\begin{align}
    1-G
    &=
    \frac{8\varphi_T^{3/2}}{5}
    \frac{\Gamma}{\bar g}
    \sqrt{\frac{2}{M}}
\simeq2.40\times10^{-4}.
\end{align}

\subsubsection*{Nuclear-spin qubits mediated by an NV-center electron}

For the two ${}^{13}\mathrm{C}$ nuclear spins, the measured hyperfine
components are \cite{supp:Taminiau2014}
\begin{equation}
    \frac{(A_{\parallel,1},A_{\perp,1})}{2\pi}
    =
    (78.2,30)~\mathrm{kHz},
    \qquad
    \frac{(A_{\parallel,2},A_{\perp,2})}{2\pi}
    =
    (32,44)~\mathrm{kHz}.
\end{equation}
Using
\begin{equation}
    \frac{\widetilde g_i}{2\pi}
    =
    \frac{1}{4}
    \sqrt{
        \left(\frac{A_{\parallel,i}}{2\pi}\right)^2
        +
        \left(\frac{A_{\perp,i}}{2\pi}\right)^2
    }
\end{equation}
gives $\bar g/2\pi=16.5~\mathrm{kHz}$. We use the conservative
electron depopulation time $T_{1}=3.24~\mathrm{ms}$. Optical
initialization into the $m_s=0$ state was performed using
$2$--$4~\mu\mathrm{s}$ laser pulses, and the shortest reported value,
$t_{\mathrm{reset}}=2~\mu\mathrm{s}$, is adopted. The estimation for the gate duration and infidelity proceeds as above.

\subsubsection*{Fluxonium qubits mediated by a transmon coupler}
The effective mediator--qubit couplings are inferred from the
state-dependent transmon transition frequencies reported in
Ref.~\cite{supp:Singh2026}. In the dressed basis, these shifts correspond
to effective longitudinal interactions
$H_{\rm int}/\hbar=\sum_i\tilde g_i Z_CZ_i$. Flipping qubit $i$
changes the coupler transition frequency by
$\delta f_i=2\tilde g_i/(2\pi)$. Averaging over the state of the
other fluxonium gives $\delta f_1=98.5~\mathrm{MHz}$ and
$\delta f_2=56.5~\mathrm{MHz}$, corresponding to
$\tilde g_1/2\pi=49.2~\mathrm{MHz}$ and
$\tilde g_2/2\pi=28.2~\mathrm{MHz}$. Their harmonic mean is
$\bar g/2\pi\simeq36.0~\mathrm{MHz}$.
 The coupler relaxation time at the operating point is approximately
$T_{1}=1.2~\mu\mathrm{s}$. A related tunable-coupler architecture
demonstrated transfer of the coupler excitation to a readout resonator
in $22~\mathrm{ns}$~\cite{supp:Chen2024}, which is used as
$t_{\mathrm{reset}}$.
\subsection*{Rb ground-hyperfine qubits mediated by a Rydberg state}

A neutral-atom realization compatible with reset-assisted control can
be constructed using a single $^{87}\mathrm{Rb}$ atom as a Rydberg
mediator and nearby $^{87}\mathrm{Rb}$ atoms as data qubits. The data
qubits remain  in the electronic ground-state hyperfine
manifold, while only the mediator is excited to a short-lived Rydberg
state. The mediator--data interaction arises from scattering of the
Rydberg electron from a ground-state atom.

Consider a data qubit encoded as
\begin{equation}
    \ket{0}=\ket{5S_{1/2},F=1},
    \qquad
    \ket{1}=\ket{5S_{1/2},F=2}.
\end{equation}
The Rydberg--ground interaction between the mediator and data atom
 may be written as
\begin{equation}
    H_{\mathrm{RF}}
    =
    n_{r,m}
    \left(
        V_1P_{1}+V_2P_{2}
    \right),
\end{equation}
where $n_{r,m}$ projects the mediator onto its Rydberg state,
$P_{F}$ projects the data atom onto ground-state hyperfine manifold
$F$, and $V_F$ is the corresponding interaction energy. Using
\begin{equation}
    n_{r,m}=\frac{1-Z_m}{2},
    \qquad
    P_{1}=\frac{1+Z}{2},
    \qquad
    P_{2}=\frac{1-Z}{2},
\end{equation}
the interaction contains the two-body term
\begin{equation}
    -\frac{V_1-V_2}{4}Z_mZ.
\end{equation}
The corresponding effective mediator--data coupling is therefore
\begin{equation}
    \frac{\widetilde g}{2\pi}
    =
    \frac{|V_1-V_2|}{4h}.
\end{equation}

MacLennan \textit{et al.} measured the binding energies of long-range
Rb$_2$ molecules formed by a $24D_J$ Rydberg atom and a
ground-state $5S_{1/2}$ atom in either hyperfine
manifold~\cite{supp:MacLennan2019}. For a shallow molecular vibrational state associated with the
$^{87}\mathrm{Rb}\,(24D_{5/2}+5S_{1/2})$ pair state, labelled
$B_1$ in Ref.~\cite{supp:MacLennan2019}, the measured binding energies are
\begin{equation}
    \frac{|V_1|}{h}=261.2~\mathrm{MHz},
    \qquad
    \frac{|V_2|}{h}=173.9~\mathrm{MHz}.
\end{equation}
Their difference is
\begin{equation}
    \frac{|V_1-V_2|}{h}=87.3~\mathrm{MHz},
\end{equation}
which gives
\begin{equation}
    \frac{\widetilde g}{2\pi}
    =
    \frac{87.3~\mathrm{MHz}}{4}
    \simeq21.8~\mathrm{MHz}.
\end{equation}
Assuming comparable interactions with the two data atoms, their
harmonic-mean coupling is estimated as
\begin{equation}
    \frac{\bar g}{2\pi}\simeq21.8~\mathrm{MHz}.
\end{equation}
These measured quantities are molecular vibrational binding energies
rather than fixed-position interaction shifts. Their conversion to
$\widetilde g_k$ should therefore be regarded as an order-of-magnitude
estimate.

Selective coupling is assumed to be implemented geometrically by
transporting the chosen pair of data atoms into the Rydberg--ground
interaction region before the gate, while keeping spectator atoms
outside the appreciable Rydberg-electron density. The selected pair can
remain in this region throughout all $M$ segments, so alternating
transport between the two data atoms is unnecessary. Optical-tweezer
experiments have demonstrated the controlled assembly of individual
Rydberg molecules comprising a Rydberg atom and a ground-state
atom~\cite{supp:Guttridge2025}.

If both selected data atoms interact simultaneously with the mediator,
the effective interaction can be written as
\begin{equation}
    H_{\mathrm{int}}=Z_m A,
    \qquad
    A=\widetilde g_1 Z_1+\widetilde g_2 Z_2 .
\end{equation}
Mediator rotations then generate the two interactions $Y_mA$ and
$Z_mA$. Their commutator is
\begin{align}
    [Y_mA,Z_mA]
    &=
    2iX_mA^2
    \nonumber\\
    &=
    2iX_m
    \left[
        \left(\widetilde g_1^2+\widetilde g_2^2\right)\mathbb I
        +2\widetilde g_1\widetilde g_2 Z_1Z_2
    \right].
\end{align}
With the mediator initialized in an eigenstate of $X_m$, the first
term contributes only a known overall phase, whereas the second
generates an entangling $Z_1Z_2$ rotation. The selected pair therefore
does not need to be switched or transported separately during the
segmented sequence; it only needs to be moved into and out of the
interaction region once per gate. In addition, coherent rotations between the mediator ground and Rydberg states,
required to change the mediator basis in the commutator sequence, can be implemented using a detuned two-photon optical drive.
Ground--Rydberg Rabi oscillations have been demonstrated for nearby
Rb $nD_{5/2}$ states~\cite{supp:Johnson2008}.

For the mediator relaxation time, we use the experimentally measured
 lifetime
\begin{equation}
    T_{1}=13.5~\mu\mathrm{s},
\end{equation}
reported for the $^{85}\mathrm{Rb}\,29D_{5/2}$ Rydberg state in
Ref.~\cite{supp:Branden2010}. This state belongs to the same $D_{5/2}$
series and is relatively close in principal quantum number to the
$24D_{5/2}$ state used for the interaction estimate. The isotope
difference, $^{85}\mathrm{Rb}$ versus $^{87}\mathrm{Rb}$, is expected to have little effect on the electronic
Rydberg-state lifetime, but this value should nevertheless be regarded
as representative.

Rapid mediator reinitialization could be implemented by stimulated
transfer directly from the $24D_{5/2}$ Rydberg state to the
short-lived $5P_{3/2}$ state, followed by spontaneous decay to the
ground-state manifold, i.e.,
\begin{equation}
    24D_{5/2}
    \longrightarrow
    5P_{3/2}
    \longrightarrow
    5S_{1/2}.
\end{equation}
The first transition is the reverse of the
$5P_{3/2}\rightarrow24D_{5/2}$ excitation used in Rb Rydberg
experiments. Ref.~\cite{supp:Piotrowicz2011} directly measured the
transition dipole moments for
$5P_{3/2}\rightarrow nD_{5/2}$ with $20<n<48$, including the
$24D_{5/2}$ state. From Fig.~6 of that
work, the reduced matrix element for $n=24$ is approximately
\begin{equation}
    \mu_{\mathrm{red}}\simeq0.068\,e a_0.
\end{equation}
For the stretched-state transition
$\ket{5P_{3/2},F=3,m_F=3}\rightarrow
 \ket{24D_{5/2},F=4,m_F=4}$, the corresponding matrix element is
\begin{equation}
    d_{24}
    =
    \sqrt{\frac{2}{3}}\,\mu_{\mathrm{red}}
    \simeq0.0555\,e a_0.
\end{equation}
Using a coupling-beam power of $50~\mathrm{mW}$ and the measured
elliptical beam waists
$w_{\mathrm{maj}}=240~\mu\mathrm{m}$ and
$w_{\mathrm{min}}=172~\mu\mathrm{m}$ gives
\begin{equation}
    \frac{\Omega_{24D,5P}}{2\pi}
    =
    \frac{d_{24}}{h}
    2\sqrt{
        \frac{P}{
        \pi w_{\mathrm{maj}}w_{\mathrm{min}}c\epsilon_0}}
    \simeq17~\mathrm{MHz}.
\end{equation}
This corresponds to a transfer time
\begin{equation}
    t_{24D\rightarrow5P}
    =
    \frac{\pi}{\Omega_{24D,5P}}
    \simeq29~\mathrm{ns}.
\end{equation}

The measured $5P_{3/2}$ lifetime is \cite{supp:Gutterres2002}
\begin{equation}
    \tau_{5P}=26.25~\mathrm{ns}.
\end{equation}
The characteristic reset time is thus
estimated as
\begin{align}
    t_{\mathrm{reset}}
    &=
    t_{24D\rightarrow5P}+\tau_{5P} 
    \simeq55~\mathrm{ns}.
\end{align}

N.B.: The interaction and reset estimates both refer to the
$24D_{5/2}$ mediator state, whereas the lifetime was measured for
$^{85}\mathrm{Rb}\,29D_{5/2}$. This use of a nearby state in the same
Rydberg series provides a reasonable order-of-magnitude lifetime
estimate but does not constitute a quantitatively self-consistent
description of one existing apparatus. A dedicated implementation would
require measuring the  Rydberg--ground interaction,
lifetime, conditional optical spectrum, and reset fidelity for the
same $^{87}\mathrm{Rb}$  $24D_{5/2}$ mediator state.
\renewcommand{\refname}{Supplementary References}

\end{document}